\documentclass[12pt,onecolumn,draftcls]{IEEEtran}
\usepackage{epsf,psfrag,amssymb,amsfonts,cite}

\newtheorem{theorem}{Theorem}
\newtheorem{example}{Example}
\newtheorem{corollary}{Corollary}
\newtheorem{lemma}{Lemma}
\newtheorem{definition}{Definition}

\newtheorem{proposition}{Proposition}

\def\psfancypar#1#2{\begingroup\def\par{\endgraf\endgroup\lineskiplimit=0pt}
               \setbox2=\hbox{\large\sc #2}
               \newdimen\tmpht \tmpht \ht2 \advance\tmpht by \baselineskip
               \font\hhuge=Times-Bold at \tmpht
               \setbox1=\hbox{{\hhuge #1}}
               \count7=\tmpht \count8=\ht1
               \divide\count8 by 1000 \divide\count7 by \count8 
               \tmpht=.001\tmpht\multiply\tmpht by \count7 
               \font\hhuge=Times-Bold at \tmpht
               \setbox1=\hbox{{\hhuge #1}}
               \noindent
                \hangindent1.05\wd1
               \hangafter=-2 {\hskip-\hangindent
               \lower1\ht1\hbox{\raise1.0\ht2\copy1}%
                \kern-0\wd1}\copy2\lineskiplimit=-1000pt}

\newcommand{\beq}{\begin{equation}}
\newcommand{\eeq}{\end{equation}}
\newcommand{\bqa}{\begin{eqnarray}}
\newcommand{\eqa}{\end{eqnarray}}
\newcommand{\bqn}{\begin{eqnarray*}}
\newcommand{\eqn}{\end{eqnarray*}}
\newcommand{\nn}{\nonumber}

\newcommand{\be}{\begin{enumerate}}
\newcommand{\ee}{\end{enumerate}}
\newcommand{\bi}{\begin{itemize}}
\newcommand{\ei}{\end{itemize}}
\newcommand{\bd}{\begin{description}}
\newcommand{\ed}{\end{description}}
\newcommand{\ba}{\begin{array}}
\newcommand{\ea}{\end{array}}
\newcommand{\bde}{\begin{definition}}
\newcommand{\ede}{\end{definition}}
\newcommand{\bex}{\begin{example}}
\newcommand{\eex}{\end{example}}

\def\boxit#1{\vbox{\hrule\hbox{\vrule\kern3pt
        \vbox{\kern3pt#1\kern3pt}\kern3pt\vrule}\hrule}}

\def\reals{ { {\rm  I \kern-0.15em R }  } }
\def\complex{ {\,{{\rm C} \kern-0.50em \raise0.20ex {  |}}\, }}

\def\0bf{{\bf 0}}
\def\1bf{{\bf 1}}
\def\2bf{{\bf 2}}
\def\3bf{{\bf 3}}
\def\4bf{{\bf 4}}
\def\5bf{{\bf 5}}
\def\6bf{{\bf 6}}
\def\7bf{{\bf 7}}
\def\8bf{{\bf 8}}
\def\9bf{{\bf 9}}

\def\ubf{{\bf u}}
\def\vbf{{\bf v}}

\def\xbf{{\bf x}}
\def\ybf{{\bf y}}

\def\xbf{{\bf x}}
\def\ybf{{\bf y}}

\def\Rbf{{\bf R}}

\def\Cmat{\mathcal{C}}

\def\Mmat{\mathcal{M}}

\def\Rmat{\mathcal{R}}

\def\Xmat{\mathcal{X}}
\def\Ymat{\mathcal{Y}}

\def\Rxx{\Rbf_{\ssstyle X\kern-.1em X}}

\let\ssstyle=\scriptscriptstyle

\def\Kout{\setbox1=\hbox{\Huge\bf K}\hbox to
1.05\wd1{\hspace{.05\wd1}
}}
\def\Sout{\setbox1=\hbox{\Huge\bf S}\hbox to 1.05\wd1{\hspace{.05\wd1}
}}

\def\scalefig#1{\epsfxsize #1\textwidth}

\date{October 4, 2009}

\title{\bf \LARGE Interference Channels with One Cognitive Transmitter}
\author{\authorblockN{Yi Cao and Biao Chen}\thanks{Y. Cao and B. Chen are with Syracuse University,
Department of Electrical Engineering and Computer Science,
Syracuse, NY 13244, email: ycao01\{bichen\}@syr.edu. This work was
supported in part by the National Science Foundation under Grants
0546491 and 0905320. The material in this paper was presented in
part at the Asilomar Conference on Signals, Systems, and
Computers, Monterery, California, November 2008, Conference on
Information Sciences and Systems, Baltimore, MD, March 2009, and
at the IEEE Globecom, Honolulu, Hawaii,  December 2009.}}
\begin{document}\maketitle
\begin{abstract}
This paper studies the problem of interference channels with one
cognitive transmitter (ICOCT) where ``cognitive'' is defined from
both the noncausal and causal perspectives. For the noncausal
ICOCT, referred to as interference channels with degraded message
sets (IC-DMS), we propose a new achievable rate region that
generalizes existing achievable rate regions for IC-DMS. In the
absence of the non-cognitive transmitter, the proposed region
coincides with Marton's region for the broadcast channel. Based on
this result, the capacity region of a class of semi-deterministic
IC-DMS is established. For the causal ICOCT, due to the complexity
of the channel model, we focus primarily on the cognitive Z
interference channel (ZIC), where the interference link from the
cognitive transmitter to the primary receiver is assumed to be
absent due to practical design considerations. Capacity bounds for
such channels in different parameter regimes are obtained and the
impact of such causal cognitive ability is carefully studied. In
particular, depending on the channel parameters, the cognitive
link may not be useful in terms of enlarging the capacity region.
An optimal corner point of the capacity region is also established
for the cognitive ZIC for a certain parameter regime.
\end{abstract}

\section{Introduction}
Cognitive radios have been proposed as an enabling technology to
address the spectrum scarcity issue. Cognitive radios are capable
of sensing their environment and adjusting their parameters and
transmission modes in real time. Therefore, they can adaptively
fill the under-utilized spaces of the wireless spectrum and
greatly increase the overall spectral efficiency.

\begin{figure}[htp]
\centerline{\leavevmode \epsfxsize=4 in \epsfysize=2 in
\epsfbox{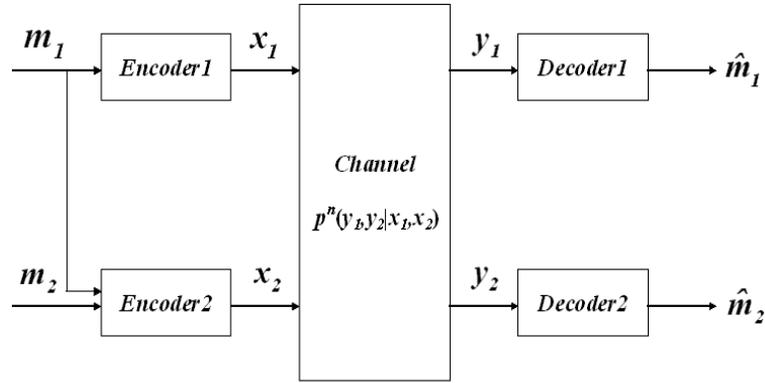}}
\caption{\label{fig:model_noncausal} Noncausal ICOCT model.}
\end{figure}

There have been recent attempts in studying the cognitive radio
channel from an information theoretic point of view
\cite{Devroye:06IT,Wu:07IT,Jovicic:06ISIT,kramer:06ITA}. There, a
cognitive radio channel is modeled as a two-user interference
channel. One of the transmitters, the so-called cognitive
transmitter, has non-causal knowledge of the other user's
transmitted messages (see Fig. \ref{fig:model_noncausal}), which
is why this model is also referred to as interference channels
with degraded message sets (IC-DMS). In \cite{Devroye:06IT}, the
authors combined Gelfand and Pinsker's
coding\cite{Gelfand&Pinsker:80PCIT} with Han and Kobayashi's
simultaneous superposition code\cite{Han&Kobayashi:81IT} to derive
an achievable rate region for the general IC-DMS. In
\cite{Wu:07IT} and \cite{Jovicic:06ISIT}, the authors derived the
capacity region for IC-DMS with weak interference. In
\cite{kramer:06ITA}, the capacity region for IC-DMS with strong
interference was determined. Those results were extended to the
Gaussian MIMO cognitive radio channels in
\cite{Sridharan&Vishwanath:07ITW}.

Recently, more general coding schemes were proposed in
\cite{Jiang&Xin:08IT} and \cite{Maric:08ETT}, which include the
results in \cite{Wu:07IT} and \cite{Jovicic:06ISIT} as special
cases. In \cite{Jiang&Xin:08IT}, the cognitive encoder uses rate
splitting and allows the other receiver to decode part of the
interference; the cognitive transmitter also cooperates by
transmitting the other user's message, and uses the Gel'fand and
Pinsker (GP) binning to cancel this known interference at its
intended receiver.  A similar approach was proposed independently
in \cite{Maric:08ETT} with the difference that they introduced the
superposition coding into their binning process, thus yielding
improvement upon \cite{Jiang&Xin:08IT}.

The IC-DMS model is interesting in the sense that it combines the
features of both interference channel and broadcast channel.
Specifically, if the cognitive user is deprived of the knowledge
about the other user's messages, it reduces to the classic
interference channel; if the primary, i.e., non-cognitive user is
absent from the channel model, it reduces to the general broadcast
channel. However, the achievable rate regions proposed in
\cite{Jiang&Xin:08IT} and \cite{Maric:08ETT} do not reduce to
Marton's region \cite{Marton:79IT} for general broadcast channels,
implying potential improvements are possible for the coding
strategy. Notice that this situation is relevant in practice if
the channel from the cognitive transmitter to the primary receiver
is superior compared with that from the primary (non-cognitive)
transmitter. Thus it is desirable to let the cognitive transmitter
take over the primary transmitter's responsibility instead of
merely serve as a cooperative transmitter. In this work, we
propose an achievable rate region for IC-DMS that generalizes the
coding schemes in \cite{Jiang&Xin:08IT,Maric:08ETT} and can also
reduce to Marton's region. The proposed new achievable region
helps establish the capacity region of a class of
semi-deterministic IC-DMS.

While the non-causal ICOCT, i.e., IC-DMS, has been extensively
studied, the causal ICOCT has been far less investigated. In the
causal scenario, the cognitive transmitter adapts its transmission
based on the causally received signals transmitted by the primary
transmitter. This causal cognitive radio model, while more
relevant and practical compared with the non-causal case, is
considerably more complex than the latter because of the noisy
feedback involved in the channel model. We remark that the causal
ICOCT is itself a special case of an even more complex model, the
so-called interference channels with generalized feedback (ICGF)
in which both transmitters are causally cognitive. Two different
achievable rate regions for ICGF were proposed in
\cite{Tuninetti:07ITA} and \cite{Cao&Chen:07ISIT} respectively,
using drastically different coding schemes. Interestingly, neither
of these two regions includes the other as a subset for the
general ICGF model, although for several extreme cases, the region
proposed in \cite{Cao&Chen:07ISIT} is shown to coincide with the
capacity region.

For the causal cognitive radio channel, we focus on the Gaussian
case and our causal ICOCT can be considered as a simplification of
ICGF, by taking away the cognitive capability from one of the
transmitters. Nonetheless, even with a single channel feedback,
the problem is still of formidable nature. In
\cite{Cao&Chen:08Asilomar}, the authors imposed a degradedness
assumption which leads to closed-form and relatively simple
capacity inner and outer bounds. In the present work, we will
instead focus on a more practical model, the so-called cognitive
ZIC for the Gaussian case where the causal ICOCT is further
simplified by \emph{taking away the interference link from the
cognitive transmitter to the primary receiver}. This Gaussian
cognitive ZIC is illustrated in Fig. \ref{fig:general_model_ZIC}.
This simplified model is largely motivated by many proposed
cognitive radio schemes that require the so-called `interference
temperature' at the primary receiver to be sufficiently low. Thus
the ZIC considered in this paper can be considered as an
approximation to such cognitive radio channels where the
interference imposed on the primary receiver by the cognitive
transmitter is largely negligible. For the cognitive ZIC, capacity
inner and outer bounds are proposed for various parameter regimes
and we demonstrate that the cognitive link may not be helpful for
certain parameter regimes, as far as the capacity region is
concerned. A corner point on the capacity region is also
established for a certain parameter regime.

This paper is organized as follows. In Section II, we consider
noncausal ICOCT and propose a new achievable rate region for
IC-DMS that generalizes existing results. The proposed region
reduces to Marton's region \cite{Marton:79IT}
 in the absence of the primary transmitter. The proposed
achievable region also allows us to establish the capacity region
for a class of semi-deterministic channels. In Section III, we
consider the causal cognitive ZIC. We propose several inner bounds
to the capacity region for different values of channel parameters.
We also introduce an outer bound to the capacity region which,
together with the inner bounds, allows us to identify a capacity
region corner point as well as parameter regimes for which the
cognitive capability does not enlarge the capacity region. The
concluding remarks are given in Section IV.

\section{Noncausal ICOCT}
\subsection{Channel Model}
A noncausal ICOCT (or IC-DMS), is a quintuple
$(\Xmat_1,\Xmat_2,p,\Ymat_1,\Ymat_2)$, where $\Xmat_1,\Xmat_2$ are
two finite input alphabet sets and $\Ymat_1,\Ymat_2$ are two
finite output alphabet sets,
$p$ is the channel transition probability $p(y_1,y_2|x_1,x_2)$. We
assume that the channels are memoryless, i.e.\bqa
p^{n}(\ybf_1,\ybf_2|\xbf_1,\xbf_2)&=&\prod_{i=1}^{n}p(y_{1i},y_{2i}|x_{1i},x_{2i})\label{eq:memoryless}\eqa
where \bqa \xbf_a=(x_{a1},\cdot\cdot\cdot,x_{an})\in\Xmat_1^{n},
\ybf_a=(y_{a1},\cdot\cdot\cdot,y_{an})\in\Ymat_1^{n}\eqa for
$a=1,2$. Let $\Mmat_1=\{1,2,\cdot\cdot\cdot,M_1\}$ and
$\Mmat_2=\{1,2,\cdot\cdot\cdot,M_2\}$ be the message sets that
sender 1 (primary transmitter) and sender 2 (cognitive
transmitter) will transmit, respectively. The cognitive
transmitter has noncausal knowledge of user 1's message, so there
are $M_1$ codewords for $\xbf_1(i)$ and $M_1\cdot M_2$ codewords
for $\xbf_2(i,j)$.
\begin{definition}
An $(M_1, M_2, n, P_e)$ code exists for the IC-DMS, if and only if
there exist two encoding functions \bqn f_1: \Mmat_1 \rightarrow
\Xmat_1^n, \hspace{.5cm} f_2: \Mmat_1\times\Mmat_2\rightarrow
\Xmat_2^n \eqn and two decoding functions \bqn g_1: \Ymat_1^n
\rightarrow \Mmat_1, \hspace{.5cm} g_2: \Ymat_2^n \rightarrow
\Mmat_2, \eqn such that $\max\{P_{e,1}^{(n)}, P_{e,2}^{(n)}\}\leq
P_e$, where $P_{e,1}^{(n)}$ and $P_{e,2}^{(n)}$ deonte the
respective average probabilties of error at decoders 1 and 2, and
are computed as \bqa P_{e,1}^{(n)}\equiv\frac{1}{M_1
M_2}\sum_{i=1}^{M_1}\sum_{j=1}^{M_2}P(\hat{m}_1\neq
i|\xbf_1(i),\xbf_2(i,j)) \\
P_{e,2}^{(n)}\equiv\frac{1}{M_1
M_2}\sum_{i=1}^{M_1}\sum_{j=1}^{M_2}P(\hat{m}_2\neq
j|\xbf_1(i),\xbf_2(i,j)) \eqa  where $\hat{m}_1$ and $\hat{m}_2$
are the decoded message index at receiver 1 and 2 respectively.
\end{definition}

\begin{definition}
A non-negative rate pair $(R_1, R_2)$ is achievable for the
IC-DMS, if for any given $0<P_e<1$ and sufficiently large $n$,
there exists a $(2^{nR_1}, 2^{nR_2}, n, P_e)$ code for the
channel. The capacity region of IC-DMS is the closure of the union
of all the achievable rate pairs $(R_1,R_2)$.
\end{definition}

\subsection{Existing Results}
The capacity region for IC-DMS with strong interference was
characterized in \cite{kramer:06ITA} and repeated in Proposition
\ref{prop:strong interference}.

\begin{proposition}\label{prop:strong interference}
\cite[Theorem 1]{kramer:06ITA} For an IC-DMS satisfying: \bqa
I(X_2;Y_2|X_1)&\leq&
I(X_2;Y_1|X_1)\\
I(X_1,X_2;Y_1)&\leq& I(X_1,X_2;Y_2) \eqa for all joint
distributions on $X_1$ and $X_2$, the capacity region $\Cmat$ is
the union of all rate pairs $(R_1,R_2)$ satisfying \bqa R_2&\leq&
I(X_2;Y_2|X_1)\\
R_1+R_2&\leq& I(X_1,X_2;Y_1) \eqa over joint distributions
$p(x_1,x_2)p(y_1,y_2|x_1,x_2)$.
\end{proposition}

The capacity rate region for IC-DMS with weak interference was
found in \cite{Wu:07IT} and \cite{Jovicic:06ISIT} as in
Proposition \ref{prop:weak interference}.
\begin{proposition}\label{prop:weak interference} \cite[Theorem 3.4]{Wu:07IT}
The capacity region for IC-DMS with weak interference is the
convex hull of all rate pairs $(R_1,R_2)$ satisfying:\bqa
R_1&\leq&
I(U,X_1;Y_1)\\
R_2&\leq& I(X_2;Y_2|U,X_1) \eqa over all probability distributions
that factor as $p(u,x_1)p(x_2|u,x_1)p(y_1,y_2|x_1,x_2)$, with the
assumption that \bqa I(X_1;Y_1)&\leq& I(X_1;Y_2)\label{event:assumption1}\\
I(U;Y_1|X_1)&\leq& I(U;Y_2|X_1)\label{event:assumption2} \eqa
\end{proposition}

In the same paper \cite{Wu:07IT}, Wu {\em et al} also proposed an
achievable region for the general IC-DMS, given in Proposition
\ref{prop:Wu's region}.

\begin{proposition}\label{prop:Wu's region}\cite[Proposition 3.1]{Wu:07IT}
The convex hull of all rate pairs $(R_1,R_2)$ satisfying \bqa
R_1&\leq& I(U,X_1;Y_1)\\
R_2&\leq& I(V;Y_2)-I(V;U,X_1) \eqa over all probability
distributions $p(x_1,x_2,u,v,y_1,y_2)$ that factor as \bqa
p(u,x_1)p(v|u,x_1)p(x_2|v,u,x_1)p(y_1,y_2|x_1,x_2) \eqa is
achievable for general IC-DMS.
\end{proposition}

More recently, Jiang and Xin proposed a general achievable rate
region \cite{Jiang&Xin:08IT} for IC-DMS, denoted by $\Rmat_{JX}$,
which includes the region proposed in \cite{Wu:07IT} and
\cite{Jovicic:06ISIT} as special cases. In order to better compare
this result with our proposed region, we provide in Proposition
\ref{prop:Jiang&Xin's region} the expressions directly derived
from the error probability analysis, i.e., before the
Fourier-Motzkin elimination.
\begin{proposition}\label{prop:Jiang&Xin's region}
The rates $R_1, R_2=R_{22}+R_{20}$ are achievable if \bqa
%R_{20}&\leq& L_{20}-I(U;W|Q)\\
%R_{22}&\leq& L_{22}-I(V;W|Q)\\
R_1&\leq& I(W;Y_1U|Q)\\
R_1+R_{20}&\leq& I(WU;Y_1|Q)\\
R_{20}&\leq& I(U;Y_2V|Q)-I(U;W|Q)\\
R_{22}&\leq& I(V;Y_2U|Q)-I(V;W|Q)\\
R_{20}+R_{22}&\leq& I(UV;Y_2|Q)+I(U;V|Q)-I(U;W|Q)-I(V;W|Q)\eqa for
some joint distribution that factors as \bqa
p(q)p(w,x_1|q)p(u|w,q)p(v|w,q)p(x_2|u,v,w,q)p(y_1,y_2|x_1,x_2)\label{eq:jiang&xin
distribution} \eqa and for which all the right-hand sides are
nonnegative.
\end{proposition}

Maric {\em et al} independently proposed an achievable rate region
for the IC-DMS \cite{Maric:08ETT}, denoted by $\Rmat_{MGKS}$ which
is given below.
\begin{proposition}\cite[Theorem 1]{Maric:08ETT} \label{prop:Maric}
Rates $R_1=R_{1a}+R_{1b}$, $R_2=R_{2a}+R_{2c}$ are achievable if
\bqa R_1&\leq&I(X_{1a},X_{1b};Y_1,U_{2c}|Q)\label{eq:maric 1}\\
R_1+R_{2c}&\leq& I(X_{1a},X_{1b},U_{2c};Y_1|Q)\label{eq:maric 2}\\
R_{1b}&\leq& I(X_{1b};Y_1,U_{2c}|X_{1a},Q)\label{eq:maric 3}\\
R_{1b}+R_{2c}&\leq& I(X_{1b},U_{2c};Y_1|X_{1a},Q)\label{eq:maric 4}\\
R_{2a}&\leq&
I(U_{2a};Y_2|U_{2c},Q)-I(U_{2a};X_{1a},X_{1b}|U_{2c},Q)\label{eq:maric 5}\\
R_2&\leq&
I(U_{2c},U_{2a};Y_2|Q)-I(U_{2c},U_{2a};X_{1a},X_{1b}|Q)\label{eq:maric
6} \eqa for some joint distribution that factors as \bqa
p(q)p(x_{1a},x_{1b},u_{2c},u_{2a},x_1,x_2|q)p(y_1,y_2|x_1,x_2)\eqa
and for which all the right-hand sides are nonnegative.
\end{proposition}

For the codebook generation, \cite{Maric:08ETT} did rate splitting
for both messages $m_1$ and $m_2$. Although $m_2$ is split into
private message $m_{2a}$ and common message $m_{2c}$, the
sub-messages from $m_1$ are both private messages, namely $m_{1a}$
and $m_{1b}$. Also, for the Gel'fand and Pinsker binning, both
$m_{2a}$ and $m_{2c}$ are encoded treating both $m_{1a}$ and
$m_{1b}$ as known interference. In other words, for the binning
part, $m_{1a}$ and $m_{1b}$ are treated as one interference.
Indeed, the same rate region as that in \cite{Maric:08ETT} can be
obtained without rate splitting for the primary user.

Without rate splitting for $m_1$, and using otherwise the same
encoding scheme and following similar error analysis, one will get
the achievable region as follows, denoted by $\Rmat_{MGKS}^{'}$
\bqa R_1&\leq&I(W;Y_1,U_{2c}|Q)\\
R_1+R_{2c}&\leq& I(W,U_{2c};Y_1|Q)\\
R_{2a}&\leq&
I(U_{2a};Y_2|U_{2c},Q)-I(U_{2a};W|U_{2c},Q)\\
R_2&\leq& I(U_{2c},U_{2a};Y_2|Q)-I(U_{2c},U_{2a};W|Q) \eqa for
some joint distribution that factors as \bqa
p(q)p(w,u_{2c},u_{2a},x_1,x_2|q)p(y_1,y_2|x_1,x_2)\eqa and for
which all the right-hand sides are nonnegative, where
$R_2=R_{2a}+R_{2c}$. We now show that the two regions, namely
$\Rmat_{MGKS}$ and $\Rmat_{MGKS}^{'}$ are identical.

First, for the region $\Rmat_{MGKS}^{'}$, if we set
$W=(X_{1a},X_{1b})$, we can get the same expressions of
(\ref{eq:maric 1})-(\ref{eq:maric 2}) and (\ref{eq:maric
5})-(\ref{eq:maric 6}), with the same joint probability
distribution. Since region $\Rmat_{MGKS}$ has two more constraints
(\ref{eq:maric 3})-(\ref{eq:maric 4}),
$\Rmat_{MGKS}\subseteq\Rmat_{MGKS}^{'}$. On the other hand, for
the region $\Rmat_{MGKS}$, if we set $X_{1b}=\phi$ and $X_{1a}=W$,
$\Rmat_{MGKS}$ is reduced to $\Rmat_{MGKS}^{'}$, so
$\Rmat_{MGKS}^{'}\subseteq\Rmat_{MGKS}$. Therefore,
$\Rmat_{MGKS}^{'}=\Rmat_{MGKS}$.

Marton in 1979 considered the general broadcast channel model and
proposed the following achievable rate region \cite{Marton:79IT}
which remains the largest to this date.
\begin{proposition}\label{prop:Marton}\cite[Theorem 2]{Marton:79IT}
Let $\Rmat_M$ be the union of rate pairs $(R_1, R_2)$ satisfying
$R_1, R_2\geq 0$ and \bqa R_1\!\!\!&\leq&\!\!\! I(WV_1;Y_1)\label{eq:marton 1}\\
R_2\!\!\!&\leq&\!\!\! I(WV_2;Y_2)\label{eq:marton 2}\\\nn
R_1+R_2\!\!\!\!&\leq&\!\!\!\! \min\{I(W;Y_1),I(W;Y_2)\}+I(V_1;Y_1|W)\\
&&+I(V_2;Y_2|W)-I(V_1;V_2|W)\label{eq:marton 3} \eqa for some
$(V_1, V_2, W)\rightarrow X\rightarrow (Y_1, Y_2)$. Then $\Rmat_M$
is achievable for the discrete memoryless broadcast channel.
\end{proposition}

\subsection{A New Inner Bound}
 Both \cite{Jiang&Xin:08IT} and \cite{Maric:08ETT} applied the
following techniques:

1) Rate splitting $R_2$ by dividing the message $m_2$ into
$m_{22}$ and $m_{20}$. Thus, the rate $R_1$ will be boosted by
letting $m_{20}$ be decoded at receiver 1  which reduces the
effective interference.

2) GP binning $m_2$ against $m_1$ so that this known interference
will be cancelled at receiver 2, boosting the rate $R_2$.

3) User 2 (cognitive transmitter) cooperates with user 1 by
transmitting message $m_1$.

However, both \cite{Jiang&Xin:08IT} and \cite{Maric:08ETT} do not
have any part of $m_1$ decoded at receiver 2. While the coding
scheme in \cite{Maric:08ETT} does involve rate splitting for
$m_1$, the two split messages, $m_{1a}$ and $m_{1b}$ in
\cite{Maric:08ETT}, are both private messages and not to be
decoded at receiver 2. This suggests potential for improvement
since GP binning is not always optimal, i.e., interference
cancellation at the receiver 2 may outperform that at the
transmitter 2 by GP binning only. For example, as observed in
\cite{Maric:08ETT}, when binning against a codebook, superposition
coding is optimal over GP binning when the interference rate is
small \cite{Maric:08ETT}. Therefore, the proposed coding scheme
further divides $m_1$ into private message $m_{11}$ and common
message $m_{10}$ and superposition encodes $m_{11}$ on top of
$m_{10}$. Additionally, since $m_{10}$ is to be completely decoded
by receiver 2, binning $m_2$ against $m_{10}$ provides no
improvements, thus, we only bin against $m_{11}$.

A second observation is that the coding schemes in
\cite{Jiang&Xin:08IT} and \cite{Maric:08ETT} let transmitter 2
help with rate 1 through coherent transmission of the noncausally
known message to receiver 1. However, if the direct link from
transmitter 1 to receiver 1 is much weaker compared with that of
the interference link from transmitter 2 to receiver 1, directly
transmitting $m_{11}$ from transmitter 2 may be suboptimal. In the
extreme case when transmitter 1 is effectively silent (due, for
example to channel conditions), transmitter 2 will serve as a
transmitter for a two user broadcast channel for which cross
binning (e.g., Marton's coding scheme) yields the largest
achievable rates. As such, the proposed coding scheme introduces
cross binning reminiscent that for the broadcast channel
\cite{ElGamal&VanderMeulen:81IT}.

The above ideas lead us to Theorem \ref{thm:new inner
bound_Fourier-Motzkin} which gives a new achievable rate region
for IC-DMS.
\begin{theorem}\label{thm:new inner bound_Fourier-Motzkin}
The rate pair $(R_1, R_2)$ is achievable for IC-DMS, if
%\bqa R_1&\leq&
%I(V_{11}U_{11}V_{20}U_{10};Y_1)-I(V_{20};U_{11}|U_{10})-I(V_{11};U_{11}|V_{20}U_{10})\label{eq:rate_FM1}\\
%R_2&\leq&
%I(V_{22}V_{20};Y_2|U_{10})-I(V_{20};U_{11}|U_{10})-I(V_{22};U_{11}|V_{20}U_{10})\label{eq:rate_FM2}\\\nn
%R_1+R_2&\leq&
%I(U_{11};Y_1V_{20}|U_{10})+I(V_{11};Y_1U_{11}|V_{20}U_{10})+I(V_{22}V_{20}U_{10};Y_2)\\
%&&-I(V_{20};U_{11}|U_{10})-I(V_{11};V_{22}|V_{20}U_{10})-I(U_{11};V_{11}V_{22}|V_{20}U_{10})\label{eq:rate_FM3}\\\nn
%R_1+R_2&\leq&
%I(V_{11}U_{11}V_{20}U_{10};Y_1)+I(V_{22};Y_2|V_{20}U_{10})\\
%&&-I(V_{20};U_{11}|U_{10})-I(V_{11};V_{22}|V_{20}U_{10})-I(U_{11};V_{11}V_{22}|V_{20}U_{10})\label{eq:rate_FM4}\\\nn
%2R_2+R_1&\leq&
%I(V_{11}U_{11}V_{20};Y_1|U_{10})+I(V_{22};Y_2|V_{20}U_{10})+I(V_{22}V_{20}U_{10};Y_2)\\
%&&-2I(V_{20};U_{11}|U_{10})-I(V_{11};V_{22}|V_{20}U_{10})-I(U_{11};V_{11}V_{22}|V_{20}U_{10})-I(V_{22};U_{11}|V_{20}U_{10})\label{eq:rate_FM5}
%\eqa
%%%%%% This following version combined some mutual information to make it look better
\bqa R_1&\leq&
I(V_{11}U_{11}V_{20}U_{10};Y_1)\label{eq:rate_FM1}\\
R_2&\leq&
I(V_{22}V_{20};Y_2|U_{10})-I(V_{22}V_{20};U_{11}|U_{10})\label{eq:rate_FM2}\\\nn
R_1+R_2&\leq&
I(V_{11}U_{11};Y_1|V_{20}U_{10})+I(V_{22}V_{20}U_{10};Y_2)\\
&&-I(V_{11};V_{22}|V_{20}U_{10})-I(U_{11};V_{22}|V_{11}V_{20}U_{10})\label{eq:rate_FM3}\\\nn
R_1+R_2&\leq&
I(V_{11}U_{11}V_{20}U_{10};Y_1)+I(V_{22};Y_2|V_{20}U_{10})\\
&&-I(V_{11};V_{22}|V_{20}U_{10})-I(U_{11};V_{22}|V_{11}V_{20}U_{10})\label{eq:rate_FM4}\\\nn
2R_2+R_1&\leq&
I(V_{11}U_{11}V_{20};Y_1|U_{10})+I(V_{22};Y_2|V_{20}U_{10})+I(V_{22}V_{20}U_{10};Y_2)\\
&&-I(V_{11};V_{22}|V_{20}U_{10})-I(U_{11};V_{22}|V_{11}V_{20}U_{10})-I(V_{22}V_{20};U_{11}|U_{10})\label{eq:rate_FM5}
\eqa for some joint distribution that factors as \bqa
p(u_{10},u_{11},v_{11},v_{20},v_{22},x_1,x_2)p(y_1,y_2|x_1,x_2)
\eqa and for which all the right-hand sides are nonnegative.
\end{theorem}
The above theorem is a direct consequence of applying the
Fourier-Motzkin elimination to the following rate region.

\begin{theorem}\label{thm:new inner bound}
Rates $R_1=R_{11}+R_{10}$ and $R_2=R_{22}+R_{20}$ are achievable
if \bqa R_{20}&\leq& L_{20}-I(V_{20};U_{11}|U_{10})\label{eq:rate 1}\\
R_{11}&\leq& L_{11}-I(V_{11};U_{11}|V_{20}U_{10})\label{eq:rate
2.1}\\
R_{22}&\leq& L_{22}-I(V_{22};U_{11}|V_{20}U_{10})\label{eq:rate
2.2}\\
R_{11}+R_{22}&\leq& L_{11}+L_{22}-I(V_{11};V_{22}|V_{20}U_{10})-I(U_{11};V_{11}V_{22}|V_{20}U_{10})\label{eq:rate 2}\\
L_{11}&\leq& I(V_{11}U_{11};Y_1|V_{20}U_{10})+I(V_{11}V_{20};U_{11}|U_{10})\label{eq:rate 3}\\
L_{11}+L_{20}&\leq& I(V_{11}U_{11}V_{20};Y_1|U_{10})+I(V_{11}V_{20};U_{11}|U_{10})\label{eq:rate 4}\\
L_{11}+L_{20}+R_{10}&\leq& I(V_{11}U_{11}V_{20}U_{10};Y_1)+I(V_{11}V_{20};U_{11}|U_{10})\label{eq:rate 5}\\
L_{22}&\leq& I(V_{22};Y_2|V_{20}U_{10})\label{eq:rate 6}\\
L_{22}+L_{20}&\leq& I(V_{22}V_{20};Y_2|U_{10})\label{eq:rate 7}\\
L_{22}+L_{20}+R_{10}&\leq& I(V_{22}V_{20}U_{10};Y_2)\label{eq:rate
8} \eqa for some joint distribution that factors as \bqa
p(u_{10},u_{11},v_{11},v_{20},v_{22},x_1,x_2)p(y_1,y_2|x_1,x_2)
\eqa and for which all the right-hand sides are nonnegative.
\end{theorem}
\begin{proof}
See Appendix.
\end{proof}

\begin{figure}[htp]
\begin{tabular}{cc}
\centerline{
\begin{psfrags}
\psfrag{W_1}[l]{$W_1$}\psfrag{W_10}[l]{$W_{10}$}
\psfrag{W_11}[l]{$W_{11}$}\psfrag{P_u10}[l]{$P_{U_{10}}$}
\psfrag{u_10}[l]{$u_{10}^n$}
\psfrag{P_u11}[l]{$P_{U_{11}|U_{10}}$}
\psfrag{u_11}[l]{$u_{11}^n$} \psfrag{f}[l]{$f(\cdot)$}
\psfrag{X_1}[l]{$X_1^n$} \scalefig{.8}\epsfbox{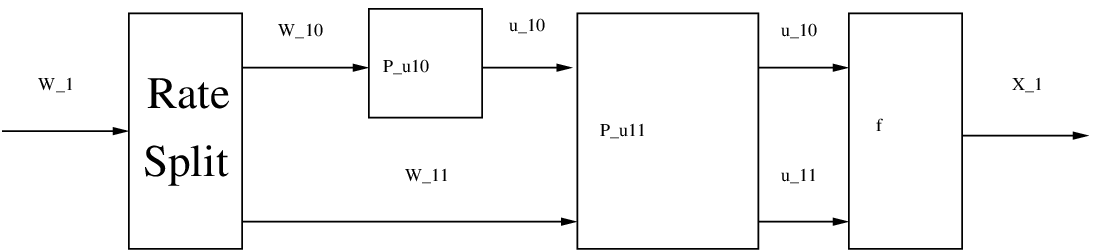}
\end{psfrags}
}\\
(a)\\
\centerline{
\begin{psfrags}
\psfrag{W_1}[l]{$W_1$}\psfrag{W_2}[l]{$W_2$}
\psfrag{W_10}[l]{$W_{10}$}\psfrag{W_11}[l]{$W_{11}$}
\psfrag{W_20}[l]{$W_{20}$}\psfrag{W_22}[l]{$W_{22}$}
\psfrag{u_10}[l]{$u_{10}^n$}
\psfrag{P_v20}[l]{$P_{V_{20}|U_{10}}$}
\psfrag{u_11}[l]{$u_{11}^n$} \psfrag{v_20}[l]{$v_{20}^n$}
\psfrag{P_v11}[l]{$P_{V_{11}|V_{20}U_{10}}$}\psfrag{P_v22}[l]{$P_{V_{22}|V_{20}U_{10}}$}
\psfrag{v_11}[l]{$v_{11}^n$} \psfrag{v_22}[l]{$v_{22}^n$}
\psfrag{g}[l]{$g(\cdot)$} \psfrag{X_2}[l]{$X_2^n$}
\scalefig{.8}\epsfbox{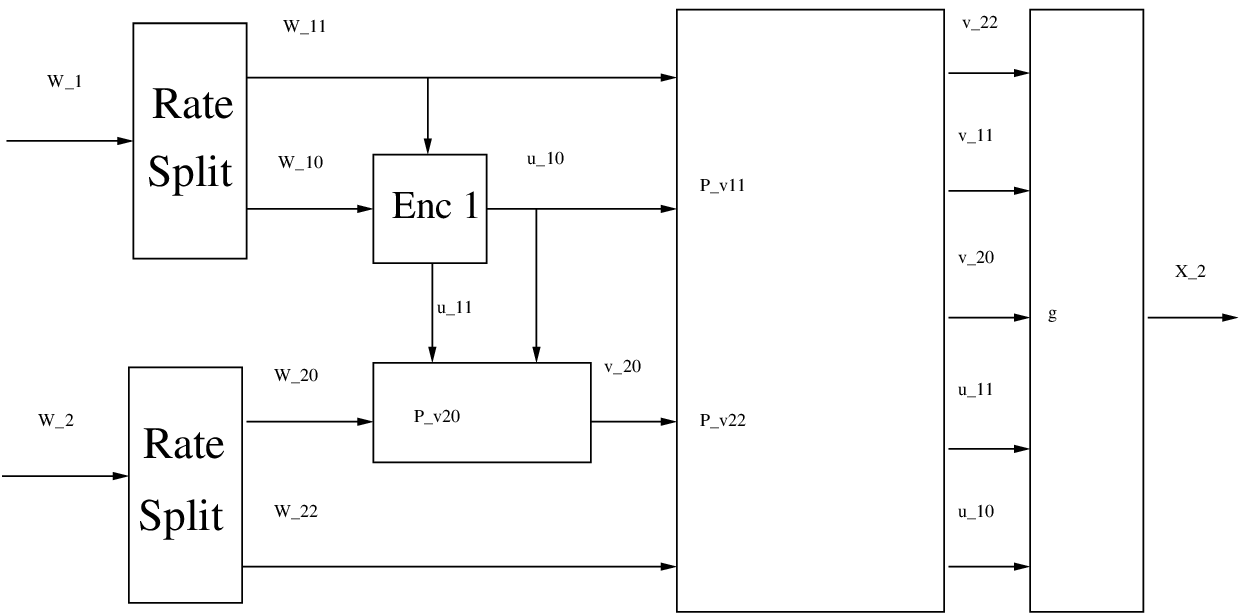}
\end{psfrags}
}\\
(b)
\end{tabular}
\caption{\label{fig:encoding} Diagram of the encoding schemes. (a)
Encoder 1 (b) Encoder 2}
\end{figure}

The encoding scheme is illustrated in Fig. \ref{fig:encoding}.
Both encoders apply rate splitting to their respective messages.
Encoder 1 encodes the two sub-messages using superposition coding.
Encoder 2, i.e., the cognitive transmitter's encoding process is
much more complex. It involves Gel'fand and Pinsker binning for
the generation of $v_{20}^n$, and the cross binning for the
generation of $v_{11}^n$ and $v_{22}^n$, as used in the general
broadcast channel. Specifically, encoder 2 first encodes message
$W_{20}$ into codeword $v_{20}^n$ superimposed on $u_{10}^n$ and
applies Gel'fand and Pinsker binning against $u_{11}^n$. Then, on
top of codeword pair $(u_{10}^n, v_{20}^n)$, it generates
$v_{11}^n$ for message $W_{11}$ and $v_{22}$ for message $W_{22}$
and applies cross binning against each other. Encoder 2 also
cooperates with the primary transmitter by transmitting codewords
$u_{11}^n$ and $u_{10}^n$.

The achievable rate region in Theorem \ref{thm:new inner bound},
denoted by $\Rmat^*$, is derived based on simultaneous decoding.
In Theorem \ref{thm:sequential inner bound}, we introduce another
region $\Rmat$ based on sequential decoding, which is a subset of
$\Rmat^*$.

\begin{theorem}\label{thm:sequential inner bound}
Rates $R_1=R_{11}+R_{10}$ and $R_2=R_{22}+R_{20}$ are achievable
if \bqa R_{20}&\leq& L_{20}-I(V_{20};U_{11}|U_{10})\label{eq:seq rate1}\\
R_{11}&\leq&
L_{11}-I(V_{11};U_{11}|V_{20}U_{10})\label{eq:seq rate2.1}\\
R_{22}&\leq& L_{22}-I(V_{22};U_{11}|V_{20}U_{10})\label{eq:seq rate2.2}\\
R_{11}+R_{22}&\leq& L_{11}+L_{22}-I(V_{11};V_{22}|V_{20}U_{10})-I(U_{11};V_{11}V_{22}|V_{20}U_{10})\label{eq:seq rate2}\\
L_{20}&\leq& \min\{I(V_{20};Y_1|U_{10}),I(V_{20};Y_2|U_{10})\}\label{eq:seq rate3}\\
R_{10}+L_{20}&\leq& \min\{I(V_{20}U_{10};Y_1),I(V_{20}U_{10};Y_2)\}\label{eq:seq rate4}\\
L_{11}&\leq& I(V_{11}U_{11};Y_1|V_{20}U_{10})+I(V_{11}V_{20};U_{11}|U_{10})\label{eq:seq rate5}\\
L_{22}&\leq& I(V_{22};Y_2|V_{20}U_{10})\label{eq:seq rate6} \eqa
for some joint distribution that factors as \bqa
p(u_{10},u_{11},v_{11},v_{20},v_{22},x_1,x_2)p(y_1,y_2|x_1,x_2)
\eqa and for which all the right-hand sides are nonnegative.
\end{theorem}
\begin{proof}
The encoding scheme is the same as that in Theorem \ref{thm:new
inner bound}, which leads to (\ref{eq:seq rate1})-(\ref{eq:seq
rate2}). The decoders first decode messages $m_{20}$ and $m_{10}$
using simultaneous decoding, which leads to (\ref{eq:seq
rate3})-(\ref{eq:seq rate4}). After subtracting out the signals
decoded in the first stage, decoder 1 proceeds to decode $m_{11}$
and decoder 2 proceeds to decode $m_{22}$, which leads to
(\ref{eq:seq rate5})-(\ref{eq:seq rate6}).
\end{proof}
\vspace{.5cm}

Both the above two regions $\Rmat^*$ and $\Rmat$ are convex.
Therefore, no convex hull operation or time sharing is necessary.
The convexity of the regions can be easily proved following the
same approach as in \cite[Lemma 5]{Csiszar&Korner:78IT}.

\subsection{Special Cases}
\subsubsection{Strong interference} In the case of strong interference, the optimal scheme is for
both user's messages to be decoded by both receivers. Thus, by
setting $V_{11}=V_{22}=U_{11}=\phi$ and $R_1=R_{10}, R_2=R_{20}$,
$\Rmat^*$ reduces to the capacity region for IC-DMS with strong
interference in Proposition \ref{prop:strong interference}.

\vspace{.5cm}

\subsubsection{Weak interference} By setting $V_{11}=V_{20}=U_{11}=\phi$,
$V_{22}=X_2$, $U_{10}=(U,X_1)$ and $R_1=R_{10}$, $R_2=R_{22}$, and
removing all the redundant conditions based on the assumption
(\ref{event:assumption1})-(\ref{event:assumption2}), $\Rmat^*$
reduces to the capacity region for IC-DMS with weak interference
in Proposition \ref{prop:weak interference}.

\vspace{.5cm}

\subsubsection{The rate region in \cite{Wu:07IT}} By setting $V_{20}=U_{10}=V_{11}=\phi$,
$U_{11}=(U,X_1)$, $V_{22}=V$ and $R_1=R_{11}=L_{11}$ and
$R_2=R_{22}$, $\Rmat^*$ reduces to Wu {\em et al}'s achievable
rate region for the general IC-DMS in Proposition \ref{prop:Wu's
region}.

\vspace{.5cm}

\subsubsection{The rate region in \cite{Devroye:06IT}} Our scheme in Theorem
\ref{thm:new inner bound} is similar to the coding scheme proposed
in \cite{Devroye:06IT} in the sense that, both users' messages are
divided into two parts: private message decoded only by the
intended receivers, and common message decoded by both receivers.
However, our scheme is different from that in \cite{Devroye:06IT}
in the following ways.
\begin{itemize}
\item User 2's codewords $\vbf_{22}$ and $\vbf_{20}$ are binned
against $\ubf_{11}$ only, while in \cite{Devroye:06IT},
$\vbf_{22}$ and $\vbf_{20}$ are binned against both $\ubf_{11}$
and $\ubf_{10}$. Since $\ubf_{10}$ is to be completely decoded by
receiver 1, binning against $\ubf_{10}$ provides no improvement.

\item The binning of $\vbf_{22}$ and $\vbf_{20}$ in
\cite{Devroye:06IT} is done independently, whereas we add
dependency between them to provide potential improvements.

\item In our scheme, the cognitive transmitter will cooperate with
the primary user by transmitting the primary user's messages,
whereas there is no cooperation between the two users in
\cite{Devroye:06IT}.

\item The coding scheme in \cite{Devroye:06IT} does not have the
codeword $\vbf_{11}$ as in our scheme. The function of $\vbf_{11}$
is to potentially cancel out the interference from the message
$m_{22}$ at receiver 1, thus boosting the rate $R_{11}$.
\end{itemize}

After applying the Fourier Motzkin elimination, for fixed joint
distribution of those random variables, the polygon of $\Rmat^*$
has constraints for four different slopes, namely, $R_1$, $R_2$,
$R_1+R_2$ and $2R_2+R_1$, while the polygon of the region in
\cite{Devroye:06IT} has one more slope constraint $2R_1+R_2$. For
fixed joint distribution, it turns out that the the bound for the
slope $2R_2+R_1$ of the region in \cite{Devroye:06IT} can be
larger than that of $\Rmat^*$, while the rest of the bounds are
smaller than $\Rmat^*$. As such, it is not easy to establish a
subset relation between these two regions analytically.

\vspace{.5cm}

\subsubsection{Jiang and Xin's region} Jiang and Xin's region
$\Rmat_{JX}$ is different from $\Rmat^*$ in that
\begin{itemize}
\item there is no rate splitting for $R_1$;

\item the binning of $\vbf_{22}$ and $\vbf_{20}$ are done
independently;

\item there is no codeword $\vbf_{11}$.
\end{itemize}
After setting $U_{10}=Q$, $U_{11}=W$, $V_{11}=\phi$, $V_{20}=U$,
$V_{22}=V$, $R_1=R_{11}=L_{11}$, and substituting all the $L_{22}$
and $L_{20}$ using (\ref{eq:rate 1})-(\ref{eq:rate 2}), $\Rmat^*$
reduces to a region, denoted by $\Rmat^{'}$, with only the
following two bounds different from $\Rmat_{JX}$: \bqa
R_{22}&\leq&
I(V;Y_2|UQ)-I(V;W|UQ)\label{eq:jiang&xin compare1}\\
R_{22}+R_{20}&\leq&
I(UV;Y_2|Q)-I(V;W|UQ)-I(U;W|Q)\label{eq:jiang&xin compare2} \eqa

The corresponding bounds in $\Rmat_{JX}$ are
\bqa R_{22}&\leq& I(V;Y_2U|Q)-I(V;W|Q)\label{eq:jiang&xin compare3}\\
R_{20}&\leq& I(U;Y_2V|Q)-I(U;W|Q)\label{eq:jiang&xin compare4}\\
R_{22}+R_{20}&\leq&
I(UV;Y_2|Q)+I(U;V|Q)-I(V;W|Q)-I(U;W|Q)\label{eq:jiang&xin
compare5} \eqa

According to the distribution (\ref{eq:jiang&xin distribution}),
for any $(V, U, W)$ in $\Rmat_{JX}$, we have \bqa
H(V|UWQ)=H(V|WQ)\label{eq:jiang&xin relation} \eqa Now, we set the
variables in $\Rmat^{'}$ as $V^*=(V,U), U^*=U, W^*=W$. Due to
(\ref{eq:jiang&xin relation}), it can be easily checked that
(\ref{eq:jiang&xin compare1}) is equal to (\ref{eq:jiang&xin
compare3}), and (\ref{eq:jiang&xin compare2}) is equal to
(\ref{eq:jiang&xin compare5}). Therefore, $\Rmat_{JX}\subseteq
\Rmat^{'}$. Hence, $\Rmat_{JX}\subseteq \Rmat^{*}$.

\vspace{.5cm}

\subsubsection{Maric \emph{et al}'s region} Maric et al's region is
different from $\Rmat^*$ in that \begin{itemize} \item no part of
$m_1$ is decoded by receiver 2; \item  there is no codeword
$\vbf_{11}$.
\end{itemize}
 By setting $U_{10}=Q$,
$U_{11}=(X_{1a},X_{1b})$, $V_{11}=\phi$, $V_{22}=U_{2a}$,
$V_{20}=U_{2c}$, $R_1=R_{11}=L_{11}$ and substituting all $L_{22}$
and $L_{20}$ by $R_{22}$ and $R_{20}$ using (\ref{eq:rate
1})-(\ref{eq:rate 2}), $\Rmat^*$ reduces to exactly (\ref{eq:maric
1})-(\ref{eq:maric 2}) and (\ref{eq:maric 5})-(\ref{eq:maric 6}).
Thus, $\Rmat^*$ includes Maric {\em et al}'s region as a subset.

\vspace{.5cm}

\subsubsection{Marton's region} In the absence of transmitter 1, IC-DMS
reduces to the general broadcast channel. However, the achievable
rate regions proposed in \cite{Jiang&Xin:08IT} and
\cite{Maric:08ETT} do not reduce to Marton's region
\cite{Marton:79IT}. This is due to the way binning is used in
\cite{Jiang&Xin:08IT} and \cite{Marton:79IT}: the binning is
always in one direction, i.e., the primary user's message is
always treated as known interference. The new ingredient in the
present work is the use of cross binning which allows us to
recover Marton's region for the broadcast channel. We now
establish that $\Rmat^*$ includes Marton's achievable region as a
subset in the absence of transmitter 1. Toward that end, it will
be convenient if we compare the region $\Rmat$ described in
Theorem \ref{thm:sequential inner bound} with Marton's region
$\Rmat_M$.

Setting $U_{11}=U_{10}=\phi$, $V_{20}=W$, $V_{11}=V_1$,
$V_{22}=V_2$ and removing redundant constraints,  the proposed region $\Rmat$ becomes \bqa R_{20}&\leq& L_{20}\label{eq:marton comp 1}\\
R_{11}+R_{22}&\leq& L_{11}+L_{22}-I(V_1;V_2|W)\\
R_{10}+L_{20}&\leq& \min\{I(W;Y_1),I(W;Y_2)\}\\
R_{11}\leq L_{11}&\leq& I(V_{1};Y_1|W)\\
R_{22}\leq L_{22}&\leq& I(V_{2};Y_2|W)\label{eq:marton comp 5}
 \eqa
Applying the Fourier-Motzkin elimination on (\ref{eq:marton comp
1})-(\ref{eq:marton comp 5}) with the definition
$R_1=R_{11}+R_{10}$ and $R_2=R_{22}+R_{20}$, $\Rmat$ becomes \bqa
R_1&\leq&
I(V_{1};Y_1|W)+ \min\{I(W;Y_1),I(W;Y_2)\}\label{eq:marton equiv 1}\\
R_2&\leq& I(V_{2};Y_2|W)+ \min\{I(W;Y_1),I(W;Y_2)\}\\
R_1+R_2&\leq& \min\{I(W;Y_1),I(W;Y_2)\}+I(V_{1};Y_1|W)+
I(V_{2};Y_2|W)-I(V_1;V_2|W)\label{eq:marton equiv 3}\eqa

The equivalence of region (\ref{eq:marton equiv
1})-(\ref{eq:marton equiv 3}) and $\Rmat_M$ was proved by Gel'fand
and Pinsker in \cite{Gelfand&Pinsker:80PIT}. Thus, we have shown
that $\Rmat$ reduces to Marton's region $\Rmat_M$ when user 1 is
absent, i.e., when the IC-DMS reduces to a broadcast channel.
Since $\Rmat\subseteq\Rmat^*$, $\Rmat^*$ includes $\Rmat_M$ as a
subset.

%Applying Fourier-Motzkin elimination to (\ref{eq:rate
%1})-(\ref{eq:rate 8}) in Theorem \ref{thm:new inner bound}, we
%obtain an alternative expression of $\Rmat^*$ in Theorem
%\ref{thm:new inner bound_Fourier-Motzkin}.

We now introduce an outer bound to the capacity region for the
general IC-DMS.
\begin{theorem}\label{thm:outer_noncausal}
An outer bound $\Rmat_o$ to the capacity region of IC-DMS is the
union of all rate pairs
$(R_1, R_2)$ satisfying \bqa R_1&\leq& I(X_1U;Y_1),\\
R_2&\leq& I(X_2;Y_2|X_1),\\
R_1+R_2&\leq& I(X_1U;Y_1)+I(X_2;Y_2|X_1U),  \eqa for some joint
distribution that factors as \bqa p(u,x_1,x_2)p(y_1,y_2|x_1,x_2).
\eqa
\end{theorem}
\begin{proof}  We start with the rate for the cognitive user.
\bqa nR_2&=&H(W_2)=H(W_2|W_1X_1^n)\label{eq:outer_noncausal_proof1}\\
&=& I(W_2;Y_2^n|W_1X_1^n)+H(W_2|Y_2^nW_1X_1^n)\label{eq:outer_noncausal_proof2}\\
&\leq& I(W_2;Y_2^n|W_1X_1^n)+n\epsilon_1\label{eq:outer_noncausal_proof3}\\
&=& \sum_{i=1}^n I(W_2;Y_{2i}|W_1X_1^nY_2^{i-1})+n\epsilon_1\label{eq:outer_noncausal_proof4}\\
&\leq& \sum_{i=1}^n
H(Y_{2i}|X_{1i})-H(Y_{2i}|W_1W_2X_1^nY_2^{i-1}X_2^n)+n\epsilon_1\label{eq:outer_noncausal_proof5}\\
&=& \sum_{i=1}^n H(Y_{2i}|X_{1i})-H(Y_{2i}|X_{1i}X_{2i})+n\epsilon_1\label{eq:outer_noncausal_proof6}\\
&=& \sum_{i=1}^n I(X_{2i};Y_{2i}|X_{1i})+n\epsilon_1
\label{eq:outer_noncausal_proof7}\eqa where
(\ref{eq:outer_noncausal_proof1}) is from the independence of the
messages, (\ref{eq:outer_noncausal_proof3}) is from Fano's
inequality, (\ref{eq:outer_noncausal_proof4}) is from the chain
rule for mutual information, (\ref{eq:outer_noncausal_proof5}) is
from conditioning does not increase entropy,
(\ref{eq:outer_noncausal_proof6}) is due to the Markov chain
$(W_1W_2X_1^nY_2^{i-1}X_2^n)\rightarrow X_{1i}X_{2i}\rightarrow
Y_{2i}$ as a result of the memoryless property of the channel.

Define the random variable $U_i=(W_1,Y_1^{i-1},Y_{2,i+1}^n)$, then
we have, for the primary user,
\bqa nR_1&=&H(W_1)\leq I(W_1;Y_1^n)+n\epsilon_2\\
&=& \sum_{i=1}^n I(W_1;Y_{1i}|Y_1^{i-1})+n\epsilon_2\\
&\leq& \sum_{i=1}^n
I(X_{1i}W_1Y_1^{i-1}Y_{2,i+1}^n;Y_{1i})+n\epsilon_2\\
&=&\sum_{i=1}^n I(X_{1i}U_i;Y_{1i})+n\epsilon_2\\
n(R_1+R_2)&=&H(W_1)+H(W_2|W_1)\leq
I(W_1;Y_1^n)+I(W_2;Y_2^n|W_1)+n\epsilon_3\\
&=& \sum_{i=1}^n
\left[I(W_1;Y_{1i}|Y_1^{i-1})+I(W_2;Y_{2i}|W_1Y_{2,i+1}^n)\right]+n\epsilon_3\\
&\leq& \sum_{i=1}^n \left[I(X_{1i}W_1Y_1^{i-1};Y_{1i})+ I(W_2;Y_{2i}|X_{1i}W_1Y_{2,i+1}^n)\right]+n\epsilon_3\label{eq:outer_noncausal_proof7}\\
&\leq& \sum_{i=1}^n
\left[I(X_{1i}W_1Y_1^{i-1};Y_{1i})+I(X_{2i}Y_1^{i-1};Y_{2i}|X_{1i}W_1Y_{2,i+1}^n)\right]+n\epsilon_3\label{eq:outer_noncausal_proof8}\\\nn
&=& \sum_{i=1}^n
\left[I(X_{1i}W_1Y_1^{i-1}Y_{2,i+1}^n;Y_{1i})-I(Y_{2,i+1}^n;Y_{1i}|X_{1i}W_1Y_1^{i_1})\right.\\
&&\left.+I(Y_1^{i-1};Y_{2i}|X_{1i}W_1Y_{2,i+1}^n)+I(X_{2i};Y_{2i}|X_{1i}W_1Y_{2,i+1}^nY_1^{i-1})\right]+n\epsilon_3\\
&=& \sum_{i=1}^n
\left[I(X_{1i}U_i;Y_{1i})+I(X_{2i};Y_{2i}|X_{1i}U_i)\right]+n\epsilon_3\label{eq:outer_noncausal_proof10}\eqa
where (\ref{eq:outer_noncausal_proof7}) is due to the
deterministic encoding at the primary user;
(\ref{eq:outer_noncausal_proof8}) is because conditioning reduces
entropy and the memoryless channel assumption;
(\ref{eq:outer_noncausal_proof10}) is due to Csiszar's identity
\cite[Lemma 7]{Csiszar&Korner:78IT}. Now, we introduce a time
sharing random variable $I$ that is independent of all other
variables and uniformly distributed over $\{1,2,...,n\}$. Define
$U=(U_I,I)$, $X_1=X_{1I}$, $X_2=X_{2I}$, $Y_1=Y_{1I}$,
$Y_2=Y_{2I}$, then the Markov chain $U\rightarrow
(X_1,X_2)\rightarrow (Y_1,Y_2)$ holds and we have \bqa nR_2&\leq&
\sum_{i=1}^n I(X_{2i};Y_{2i}|X_{1i})+n\epsilon_1\\
&=& n I(X_{2I};Y_{2I}|X_{1I}I)+n\epsilon_1\\
&=& n[H(Y_{2I}|X_{1I}I)-H(Y_{2I}|X_{1I}X_{2I}I)]+n\epsilon_1\\
&\leq& n[H(Y_{2I}|X_{1I})-H(Y_{2I}|X_{1I}X_{2I})]+n\epsilon_1\\
&=& n I(X_{2I};Y_{2I}|X_{1I})+n\epsilon_1\\
&=& n I(X_2;Y_2|X_1)+n\epsilon_1\\
nR_1 &\leq& \sum_{i=1}^n
I(X_{1i}U_i;Y_{1i})+n\epsilon_2\\
&=& n I(X_{1I}U_I;Y_{1I}|I)+n\epsilon_2\\
&\leq& n I(X_{1I}U_II;Y_{1I})+n\epsilon_2\\
&=& n I(X_1U;Y_1)+n\epsilon_2\\
n(R_1+R_2)&\leq& \sum_{i=1}^n
I(X_{1i}U_i;Y_{1i})+I(W_2;Y_{2i}|X_{1i}U_i)+n\epsilon_3\\
&=& n [I(X_{1I}U_I;Y_{1I}|I)+I(X_{2I};Y_{2I}|X_{1I}U_II)]+n\epsilon_3\\
&\leq& n [I(X_{1I}U_II;Y_{1I})+I(X_{2I};Y_{2I}|X_{1I}U_II)]+n\epsilon_3\\
&=& n [I(X_1U;Y_1)+I(X_2;Y_2|X_1U)]+n\epsilon_3\eqa
\end{proof}

\subsection{Semi-deterministic IC-DMS}
We can now consider the semi-deterministic IC-DMS with the
deterministic component for receiver 2. More specifically, the
received signal at receiver 2 is a deterministic function of the
input signal $X_1$ and $X_2$, i.e., $Y_2=h(X_1,X_2)$. In other
words, the channel matrix from the input $(X_1,X_2)$ to the output
$Y_2$ has $0$ or $1$ as its entries.

For this semi-deterministic IC-DMS, if we impose an additional
constraint $I(X_1;Y_1)\leq I(X_1;Y_2)$, we are able to find the
capacity region using the inner bound in Theorem \ref{thm:new
inner bound_Fourier-Motzkin} and the outer bound in Theorem
\ref{thm:outer_noncausal}, as given in the following theorem.

\begin{theorem}\label{thm:capacity_semideterm}
For a semi-deterministic IC-DMS, if $I(X_1;Y_1)\leq I(X_1;Y_2)$
for any input distributions, the capacity region is the union of
all rate pairs $(R_1, R_2)$
satisfying \bqa R_1&\leq& I(X_1U;Y_1)\label{eq:capacity_semideterm1}\\
R_2&\leq& H(Y_2|X_1)\label{eq:capacity_semideterm2}\\
R_1+R_2&\leq&
I(X_1U;Y_1)+H(Y_2|X_1U)\label{eq:capacity_semideterm3} \eqa for
some joint distributions that factor as \bqa
p(u,x_1,x_2)p(y_1,y_2|x_1,x_2)\label{eq:Markov_U}. \eqa
\end{theorem}
\begin{proof}
Converse: By definition, we have \bqa
I(X_2;Y_2|X_1)=H(Y_2|X_1)\label{eq:semi_determ1}\\
I(X_2;Y_2|X_1U)=H(Y_2|X_1U)\label{eq:semi_determ2} \eqa Plug
(\ref{eq:semi_determ1})-(\ref{eq:semi_determ2}) into Theorem
\ref{thm:outer_noncausal}, we get
(\ref{eq:capacity_semideterm1})-(\ref{eq:capacity_semideterm3}).

Achievability: In the inequalities
(\ref{eq:rate_FM1})-(\ref{eq:rate_FM5}) in Theorem \ref{thm:new
inner bound_Fourier-Motzkin}, let $U_{10}=X_1$, $V_{11}=U$,
$V_{22}=X_2$, $U_{11}=V_{20}=\phi$. Then, (\ref{eq:rate_FM1}) and
(\ref{eq:rate_FM2}) reduce to
(\ref{eq:capacity_semideterm1})-(\ref{eq:capacity_semideterm2})
directly. Also, (\ref{eq:rate_FM4}) reduces to \bqa R_1+R_2\leq
I(X_1U;Y_1)+I(X_2;Y_2|X_1)-I(U;X_2|X_1).
\label{eq:semideterm_proof1}\eqa Due to the Markov chain
(\ref{eq:Markov_U}), \bqa
I(X_2;Y_2|X_1)&=&H(Y_2|X_1)-H(Y_2|X_1X_2)\\
&=&H(Y_2|X_1)-H(Y_2|X_1X_2U)\\
&=&I(Y_2;X_2U|X_1)\\
&=&I(Y_2;U|X_1)+I(Y_2;X_2|UX_1)\label{eq:semideterm_proof2} \eqa
Also, due to the semi-deterministic channel, we have \bqa
I(Y_2;U|X_1)&=&H(U|X_1)-H(U|X_1Y_2)\\
&=&H(U|X_1)-H(U|X_1X_2)\\
&=&I(U;X_2|X_1)\label{eq:semideterm_proof3} \eqa Combine
(\ref{eq:semideterm_proof2}) and (\ref{eq:semideterm_proof3}),
(\ref{eq:semideterm_proof1}) becomes \bqa R_1+R_2\leq
I(X_1U;Y_1)+I(X_2;Y_2|X_1U)=I(X_1U;Y_1)+H(Y_2|X_1U), \eqa hence
(\ref{eq:capacity_semideterm3}) is achieved. Due to the additional
constraint $I(X_1;Y_1)\leq I(X_1;Y_2)$, (\ref{eq:rate_FM3}) and
(\ref{eq:rate_FM5}) become redundant. This completes the proof for
the achievability.
\end{proof}
The capacity region in Theorem \ref{thm:capacity_semideterm}
indicates that for semi-deterministic IC-DMS with constraint
$I(X_1;Y_1)\leq I(X_1;Y_2)$, the optimal transmission scheme is
for the primary user's messages $m_1$ to be decoded by both
receivers while the cognitive user encodes $m_2$ on top of $m_1$
using superposition coding.

\section{The Cognitive ZIC}
In this section, we study ICOCT with causal cooperation. As
explained in Section I, our focus will be on the special case of
Gaussian ZIC where the interference link between the cognitive
transmitter and the primary receiver is absent.
\subsection{Channel Model}
\begin{figure}[htp]
\centerline{\leavevmode \epsfxsize=4 in \epsfysize=2 in
\epsfbox{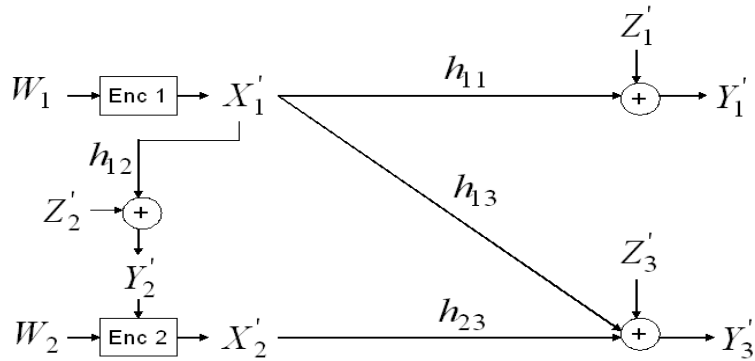}}
\caption{\label{fig:general_model_ZIC} The general model of the
cognitive ZIC.}
\end{figure}
The (causal) cognitive ZIC is illustrated in
Fig.\ref{fig:general_model_ZIC}. User 1 has message
$W_1\in\{1,2,\cdot\cdot\cdot,2^{nR_1}\}$ to be transmitted to
receiver 1 ($Y_1^{'}$), and user 2 has message
$W_2\in\{1,2,\cdot\cdot\cdot,2^{nR_2}\}$ for receiver 2
($Y_3^{'}$). In addition, user 2 can listen to the transmitted
signal from user 1 through a noisy channel ($Y_2^{'}$). Thus, the
channel model
is given by \bqa Y_1^{'}&=&h_{11}X_1^{'}+Z_1^{'}\label{eq:ZIC_model1}\\
Y_2^{'}&=&h_{12}X_1^{'}+Z_2^{'}\label{eq:echo_cancel}\\
Y_3^{'}&=&h_{13}X_1^{'}+h_{23}X_2^{'}+Z_3^{'}\label{eq:ZIC_model3}
 \eqa
where $h_{11},h_{12},h_{13}$ and $h_{23}$ are fixed real positive
numbers. $Z_1{'}\sim N(0,N_1)$, $Z_2{'}\sim N(0,N_2)$ and
$Z_3{'}\sim N(0,N_3)$ are independent Gaussian random variables.
The average power constraints of the input signals are \bqa
\frac{1}{n}\sum_{i=1}^n (x_{ti}^{'})^2\leq P_t^{'} \eqa where
$t=1,2$. From (\ref{eq:echo_cancel}), we have assumed implicitly
perfect echo cancellation.

\begin{lemma}
Any cognitive ZIC described by
(\ref{eq:ZIC_model1})-(\ref{eq:ZIC_model3}) is equivalent, in its
capacity region, to the following cognitive ZIC in standard form
\bqa Y_1&=&X_1+Z_1\label{eq:model_standard1}\\
Y_2&=&KX_1+Z_2\\
Y_3&=&bX_1+X_2+Z_3 \label{eq:model_standard3}\eqa
 where $Z_1$, $Z_2$ and $Z_3$ are
independent zero mean, unit variance Gaussian variables, and
$X_1$, $X_2$ are subject to respective power constraints $P_1$ and
$P_2$. $K$ and $b$ are deterministic real numbers with $0\leq
K<\infty$, $0\leq b<\infty$.
\end{lemma}
The proof is through a simple scaling transformation similar to
that of \cite{Carleial:75IT}, hence omitted.

\vspace{.5cm}

\begin{figure}[htp]
\centerline{\leavevmode \epsfxsize=4 in \epsfysize=2 in
\epsfbox{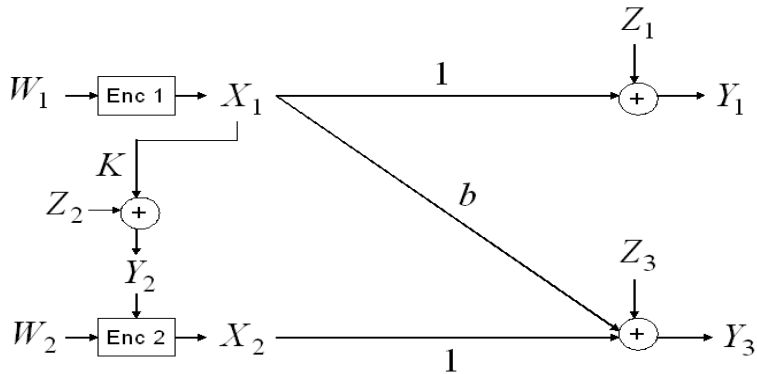}} \caption{\label{fig:model_ZIC} The
standard model of the cognitive ZIC.}
\end{figure}
The encoding functions $f_1$ and $f_2$ for users 1 and 2 are
respectively: \bqa \xbf_1&=& f_1(W_1)\\
x_{2i}&=&
f_2(W_2,y_{21},\cdot\cdot\cdot,y_{2,i-1})\label{eq:encoding
function 2}   \eqa for $i=1,2,\cdot\cdot\cdot,n$. We only consider
deterministic encoders, as nondeterministic encoders do not
enlarge the capacity region (See, e.g., \cite[Appendix
D]{Willems:85IT}).

\subsection{Capacity Lower Bounds} The cognitive ZIC includes
the following two extreme cases: the classic ZIC (corresponding to
$K=0$) and the ZIC with degraded message sets ($K=\infty$). To
simplify our notation, we define \bqa \gamma(x)\triangleq
\frac{1}{2}\log(1+x). \eqa For the classic ZIC, when $b\geq 1$,
the capacity region, denoted by $\Rmat_1$, is given below
\bqa R_1&\leq& \gamma(P_1)\triangleq C_1\\
R_2&\leq& \gamma(P_2)\triangleq C_2\\
R_1+R_2&\leq& \gamma(b^2P_1+P_2) \eqa Apparently, $\Rmat_1$ is an
inner bound to the capacity region for the corresponding cognitive
ZIC. When $b<1$, we do not know the whole capacity region, but the
sum rate capacity is known to be: \bqa R_1+R_2\leq
\gamma(P_1)+\gamma\left(\frac{P_2}{1+b^2P_1}\right) \eqa which is
achieved when user 1 is transmitting at its maximum rate, and user
2 is transmitting at a rate such that its message can be decoded
at receiver 2 by treating user 1's signal as noise.

On the other extreme, for the ZIC with degraded message sets,
where user 2 has {\em a priori} knowledge of user 1's message, the
capacity region, denoted by $\Rmat_2$, is the following rectangle,
for all $b\geq 0$: \bqa
R_1&\leq& C_1\label{eq:outer1}\\
R_2&\leq& C_2\label{eq:outer2} \eqa This is because for the
Gaussian channel considered in this paper, user 2 can dirty paper
code its own message treating user 1's signal as known
interference \cite{Costa:83IT}. Therefore, this is equivalent to
two parallel interference free channels. $\Rmat_2$ serves as a
natural outer bound to the capacity region of the cognitive ZIC.

Little is known for the cognitive ZIC besides of the two extreme
cases. The difficulty for the case with finite $K$ is that it is
not clear what is the optimal way to utilize the channel feedback
at transmitter 2. Any information overheard through the cognitive
link pertains only to message $W_1$, yet, unlike the noncausal
case described in Section II, the absence of link from $X_2$ to
$Y_1$ implies that existence of the cognitive link can not
directly benefit the rate $R_1$ through cooperative transmission.
On the other hand, since $X_1$ interferes receiver $Y_3$, encoding
schemes should explore the potential of facilitating interference
cancellation for the secondary user using the cognitive link. In
the following, we describe several cases where we can obtain
closed form capacity bounds for the cognitive ZIC and discuss
their implication in terms of the impact of the ``cognitive
capability" on the capacity.

\subsubsection{$b^2\geq 1+P_2$} In the absence of the cognitive
link, this reduces to the ZIC with very strong interference. The
capacity region of such ZIC coincides with the outer bound
$\Rmat_2$ for the cognitive ZIC, suggesting that $\Rmat_2$ is
indeed the capacity region. Notice this is the case where there is
no need to utilize the channel feedback $Y_2$ at user 2, as far as
the capacity region is concerned.

\subsubsection{$1\leq b^2<1+P_2$} This is a very interesting case.
In the absence of the cognitive link, the capacity region is the
pentagon described by $\Rmat_1$. On the other hand, for ZIC with
degraded message sets, the capacity is a rectangle, $\Rmat_2$. The
capacity region for the cognitive ZIC with $1\leq b^2<1+P_2$
should be between these two regions. We define another region
$\Rmat_3$ as the union of all nonnegative rate pairs $(R_1,R_2)$
such that:\bqa R_1\!\!&\leq&\!\!
\gamma(K^2\alpha P_1)\\
R_1\!\!&\leq&\!\! \gamma(P_1)\\
R_1\!\!&\leq&\!\!
\gamma(K^2(\alpha-\beta)P_1)+\gamma\left(\frac{b^2\beta
P_1}{1+b^2(1-\beta)P_1+P_2}\right)\\
R_1\!\!&\leq&\!\! \gamma((1-\beta)P_1)+\gamma\left(\frac{b^2\beta
P_1}{1+b^2(1-\beta)P_1+P_2}\right)\\
R_2\!\!&\leq&\!\!\gamma\left(\frac{P_2}{1+b^2(\alpha-\beta)P_1}\right)
 \eqa
where $0\leq\beta\leq\alpha\leq 1$.

\begin{theorem}
The the convex hull of the union of $\Rmat_1$ and $\Rmat_3$ is
achievable for the cognitive ZIC.
\end{theorem}
\begin{proof}
We only need to prove the achievability of $\Rmat_3$. User 1
splits the message $W_1$ into two parts: the common message
$W_{12}$, which is to be decoded by all the receivers $Y_1$, $Y_2$
and $Y_3$; the private message $W_{11}$, which is only to be
decoded by receivers $Y_1$ and $Y_2$. User 1 randomly places the
codewords of $W_{11}$ into $2^{nR_0}$ cells , indexed by $S_{11}$,
and transmits $W_{11}$ using the superposition block Markov
encoding \cite{Cover&Elgamal:79IT}, as shown in
Fig.\ref{fig:transmission scheme}.

\begin{figure}[htp]
\centerline{\leavevmode \epsfxsize=6.4 in \epsfysize=1.2 in
\epsfbox{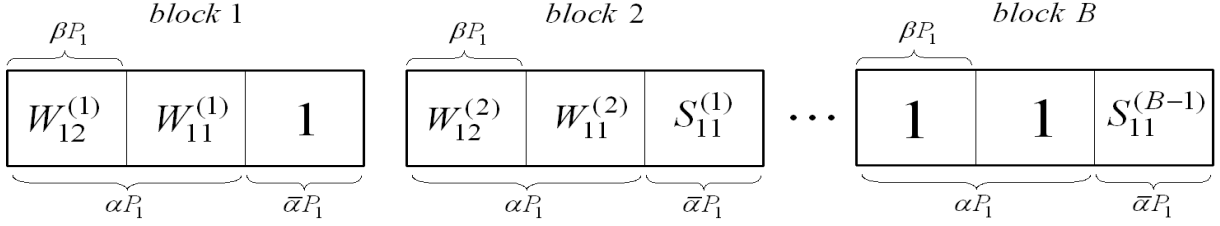}} \caption{\label{fig:transmission
scheme} The transmission scheme for $W_1$.}
\end{figure}
The whole transmission proceeds across $B$ blocks. In block $i$,
the codeword $\xbf_1^{(i)}$ includes the new message of
$W_{11}^{(i)}$ and $W_{12}^{(i)}$ for the current block and the
``cell index" $S_{11}^{(i-1)}$ of the $W_{11}^{(i-1)}$ in block
$i-1$. User 2 always decodes the new message of $W_{11}^{(i)}$ and
$W_{12}^{(i)}$ at the end of block $i$ and dirty paper code
$W_2^{(i)}$ treating the previous block's cell index
$S_{11}^{(i-1)}$ as known interference. At the end of block $i$,
receiver $Y_3$ first decodes $W_{12}^{(i)}$, subtracts it out, and
then dirty paper decodes $W_2^{(i)}$, treating the other part of
the new message, $W_{11}^{(i)}$, as noise. Therefore, the
interference corresponding to $S_{11}$ is mitigated through dirty
paper coding, boosting the rate of $W_2$.

In the first block, user 1 transmits a constant cell index
$S_{11}=1$, since there is no message $W_{11}$ in the previous
block; in the last block, user 1 transmits constant new message
$W_{11}=1$ and $W_{12}=1$. In all other blocks, user 1 allocates
$\beta P_1$ for $W_{12}$, $0\leq\beta\leq 1$; allocates
$(\alpha-\beta)P_1$ for $W_{11}$, $\beta\leq\alpha\leq 1$, and the
remaining power $(1-\alpha)P_1$ for $S_{11}$. Since $Y_2$ always
knows $S_{11}$ transmitted in the current block, the rate
constraints at $Y_2$ are \bqa R_{11}&\leq&
\gamma(K^2(\alpha-\beta)P_1)\label{eq:ZIC_rate1}\\
R_{12}&\leq& \gamma(K^2\beta P_1)\\
R_{11}+R_{12}&\leq& \gamma(K^2\alpha P_1) \eqa Receiver $Y_1$ also
decodes $W_{12}$ first, subtracts it out, then decodes $S_{11}$.
The corresponding rate constraints for reliable decoding are
respectively \bqa R_{12}&\leq& \gamma\left(\frac{\beta
P_1}{(1-\beta)P_1+1}\right)\\
R_0&\leq&
\gamma\left(\frac{(1-\alpha)P_1}{1+(\alpha-\beta)P_1)}\right).\label{eq:R_11(1)}
 \eqa

After this, $Y_1$ intersects the codewords of $W_{11}$ in bin
$S_{11}$ with the ambiguity set \cite{Cover&Elgamal:79IT} of
$W_{11}$ in the previous block to find the unique codeword
$W_{11}$. The uniqueness is guaranteed if \bqa R_{11}\leq
\gamma((\alpha-\beta)P_1)+R_0\label{eq:R_11(2)} \eqa Combine
(\ref{eq:R_11(1)}) and (\ref{eq:R_11(2)}), we have \bqa
R_{11}&\leq& \gamma((1-\beta)P_1) \eqa The rate constraints at
$Y_3$ are \bqa R_{12}&\leq& \gamma\left(\frac{b^2\beta
P_1}{1+b^2(1-\beta)P_1+P_2}\right)\\
R_2 &\leq&
\gamma\left(\frac{P_2}{1+b^2(\alpha-\beta)P_1}\right)\label{eq:ZIC_rate9}
\eqa which are the results of sequentially decoding $W_{12}$ and
$W_2$. For $R_1=R_{11}+R_{12}$, apply Fourier-Motzkin elimination
on (\ref{eq:ZIC_rate1})-(\ref{eq:ZIC_rate9}), we obtain the region
$\Rmat_3$.
\end{proof}
Notice that while the encoding scheme mimics that for the relay
channel \cite{Cover&Elgamal:79IT}, this is for a different
purpose. The reason is that the new message in each block is
required to be decoded by $Y_2$ with a maximum rate
$\gamma(K^2\alpha P_1)$. In the case of $K^2\alpha>1$, $Y_1$ will
not be able to decode this new message; the cell index transmitted
in the next block is to ensure reliable decoding of the message by
$Y_1$. Thus decoding of $W_1$ at $Y_1$ is accomplished in two
steps, much like the way decoding is done at the receiver for the
classic three node relay channel: in block $i-1$, the received
signal $Y_1$ allows the partition of message indices into
$\gamma((\alpha-\beta)P_1)$ subsets, thereby allowing the
construction of the ambiguity set; in block $i$ the cell index can
be reliably decoded if (\ref{eq:R_11(1)}) is satisfied. Uniqueness
of the message index as the intersection of the ambiguity set and
the cell is guaranteed if (\ref{eq:R_11(2)}) is satisfied. In
addition to facilitating the decoding of $W_1$ at $Y_1$, the cell
index itself allows interference cancellation for the cognitive
transceiver pair through dirty paper coding and decoding as it is
known at the beginning of the each block for transmitter 2.

\begin{figure}[htp]
\begin{tabular}{cc}
\leavevmode \epsfxsize=3.2in \epsfysize=2.4in \epsfbox{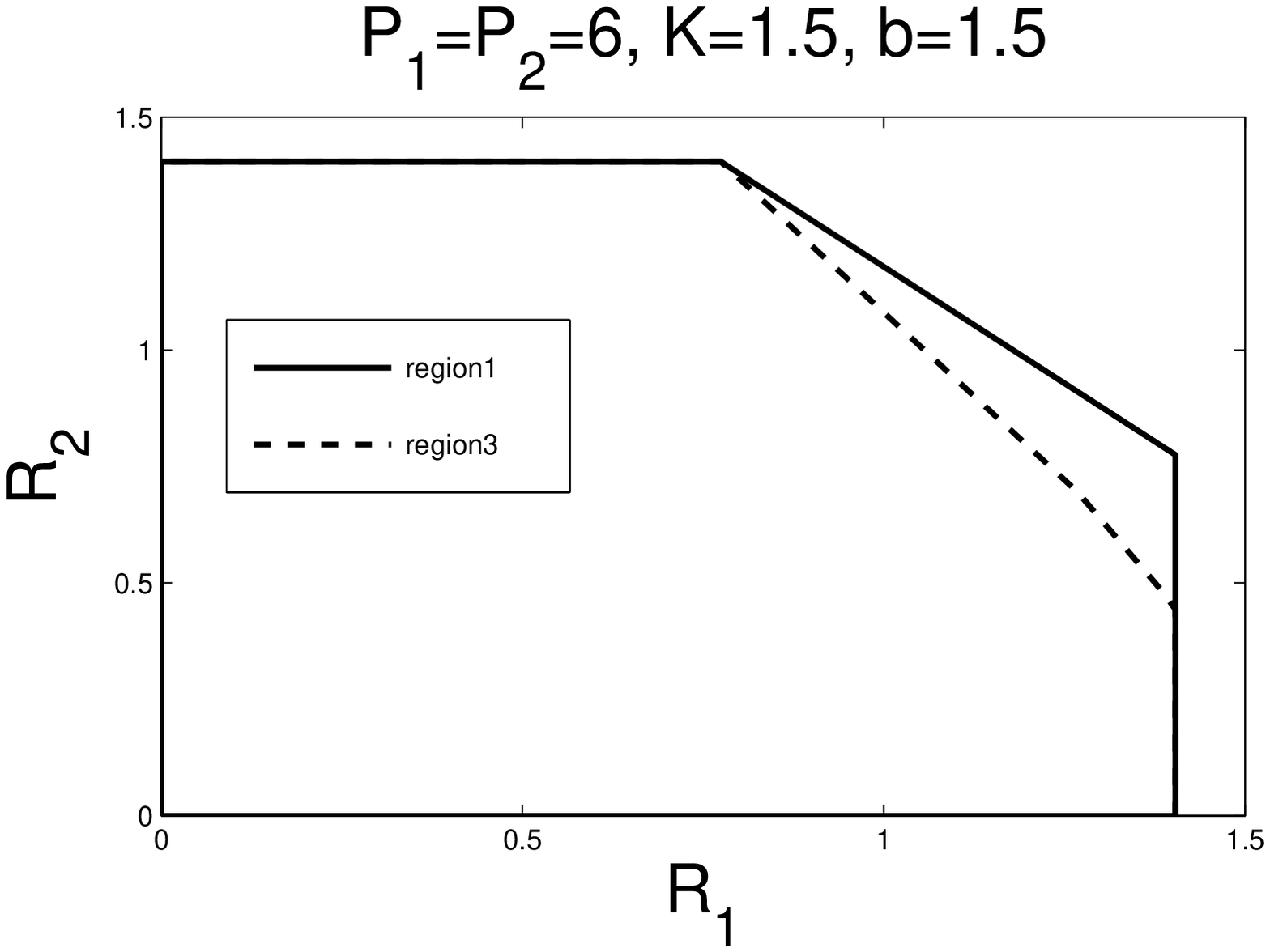} &
\leavevmode \epsfxsize=3.2in
\epsfysize=2.4in \epsfbox{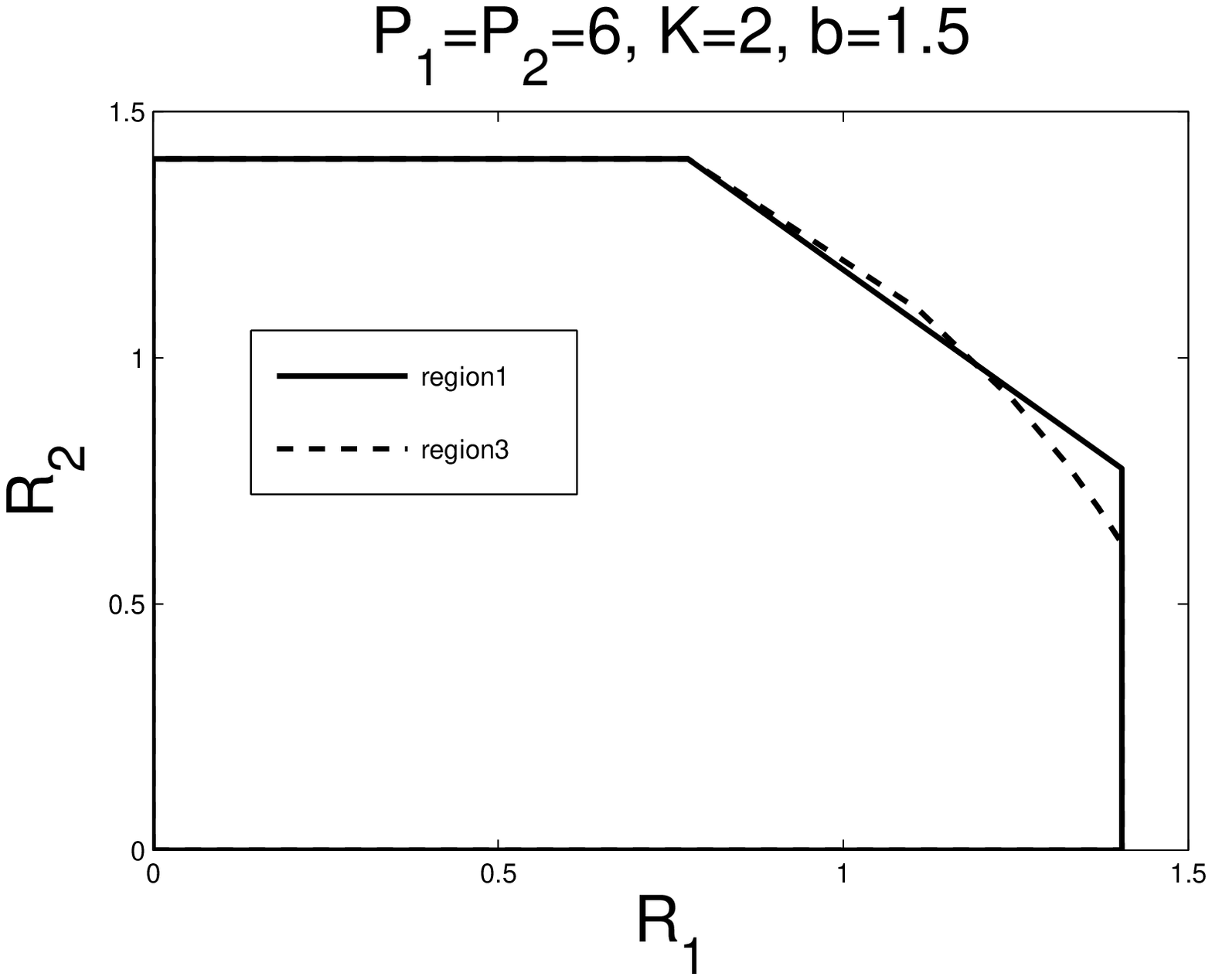}\\
(a) & (b)\\
\leavevmode \epsfxsize=3.2in \epsfysize=2.4in \epsfbox{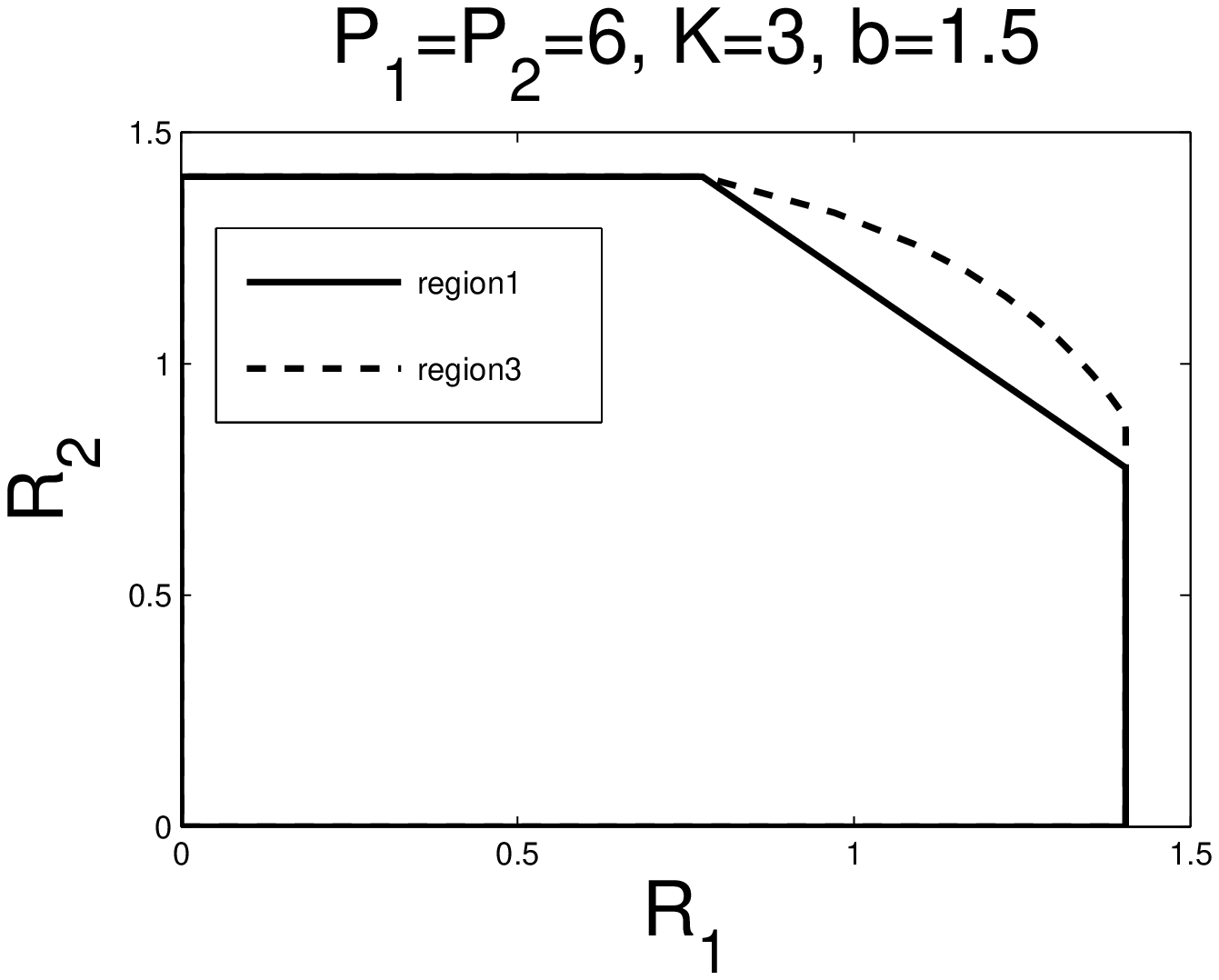}\\
(c)
\end{tabular}
\caption{\label{fig:1-3} (a) $P_1=P_2=6, K=1.5, b=1.5$. (b)
$P_1=P_2=6, K=2, b=1.5$. (c) $P_1=P_2=6, K=3, b=1.5$.}
\end{figure}
Some numerical examples are given in Fig.\ref{fig:1-3}. We see
from Fig.\ref{fig:1-3} that with different values of $K$, the
subset relation between $\Rmat_1$ and $\Rmat_3$ varies. To find
out their precise relation, we only need to consider the corner
points of the two regions as $\Rmat_1$ is a convex hull of its
corner points whereas $\Rmat_3$ is a convex region.
 For region $\Rmat_1$, when $R_1=C_1$,\bqa
R_2=\frac{1}{2}\log\left(\frac{1+b^2P_1+P_2}{1+P_1}\right).\eqa
For region $\Rmat_3$, when $R_1=C_1$, \bqa
R_2=\frac{1}{2}\log\left(1+\frac{P_2}{1+b^2P_1/K^2}\right).\eqa In
order for $\Rmat_1$ to be a subset of $\Rmat_3$, we need \bqa
\frac{1+b^2P_1+P_2}{1+P_1}\leq 1+\frac{P_2}{1+b^2P_1/K^2}\eqa
which yields \bqa K^2\geq b^2\cdot\frac{(b^2-1)P_1+P_2}{1+P_2-b^2}
\eqa That is, when $K$ is large enough, $\Rmat_3$ dominates
$\Rmat_1$, i.e., the coding scheme where receiver 2 needs to
decode both users' messages, which is capacity achieving for the
classic ZIC with strong interference, is no longer optimal here.
However, when $K$ is small enough, especially when $K$ is near
$1$, $\Rmat_1$ will dominate $\Rmat_3$, thus the cognitive link
between the two users appears to yield no rate gain for the
proposed encoding scheme.

\begin{figure}[htp]
\begin{tabular}{cc}
\leavevmode \epsfxsize=3.2in \epsfysize=2.4in \epsfbox{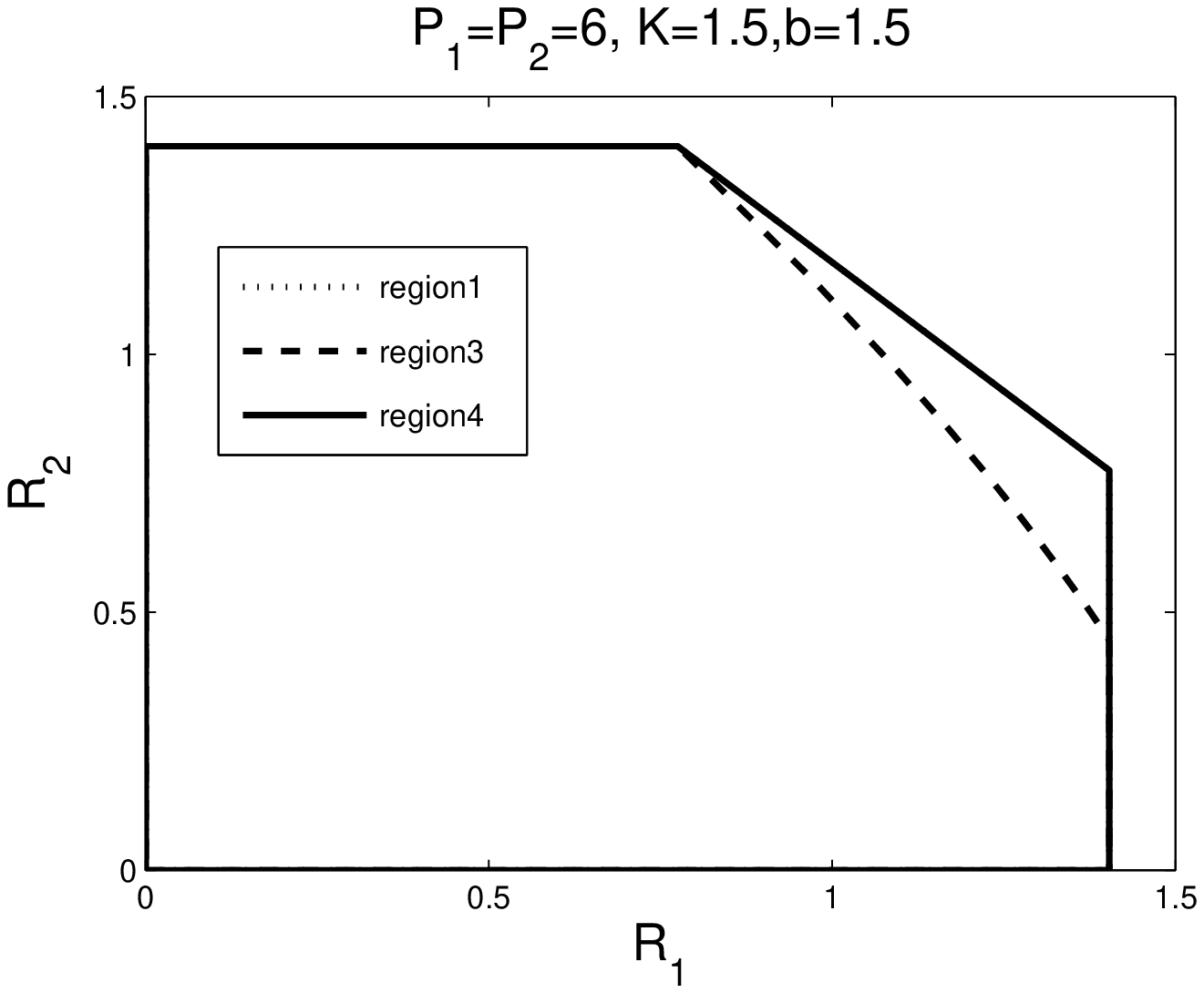} &
\leavevmode \epsfxsize=3.2in
\epsfysize=2.4in \epsfbox{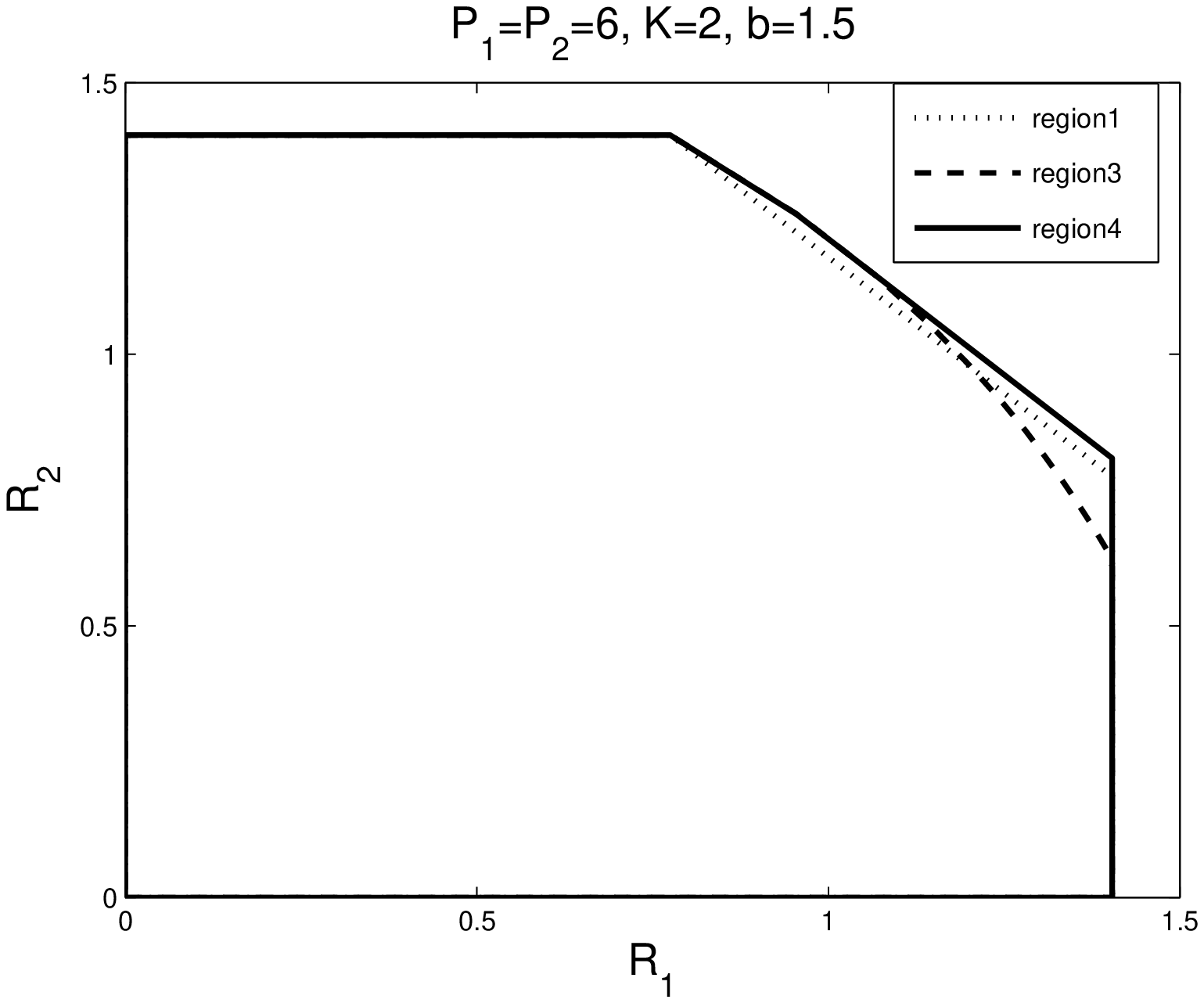}\\
(a) & (b)\\
\leavevmode \epsfxsize=3.2in \epsfysize=2.4in \epsfbox{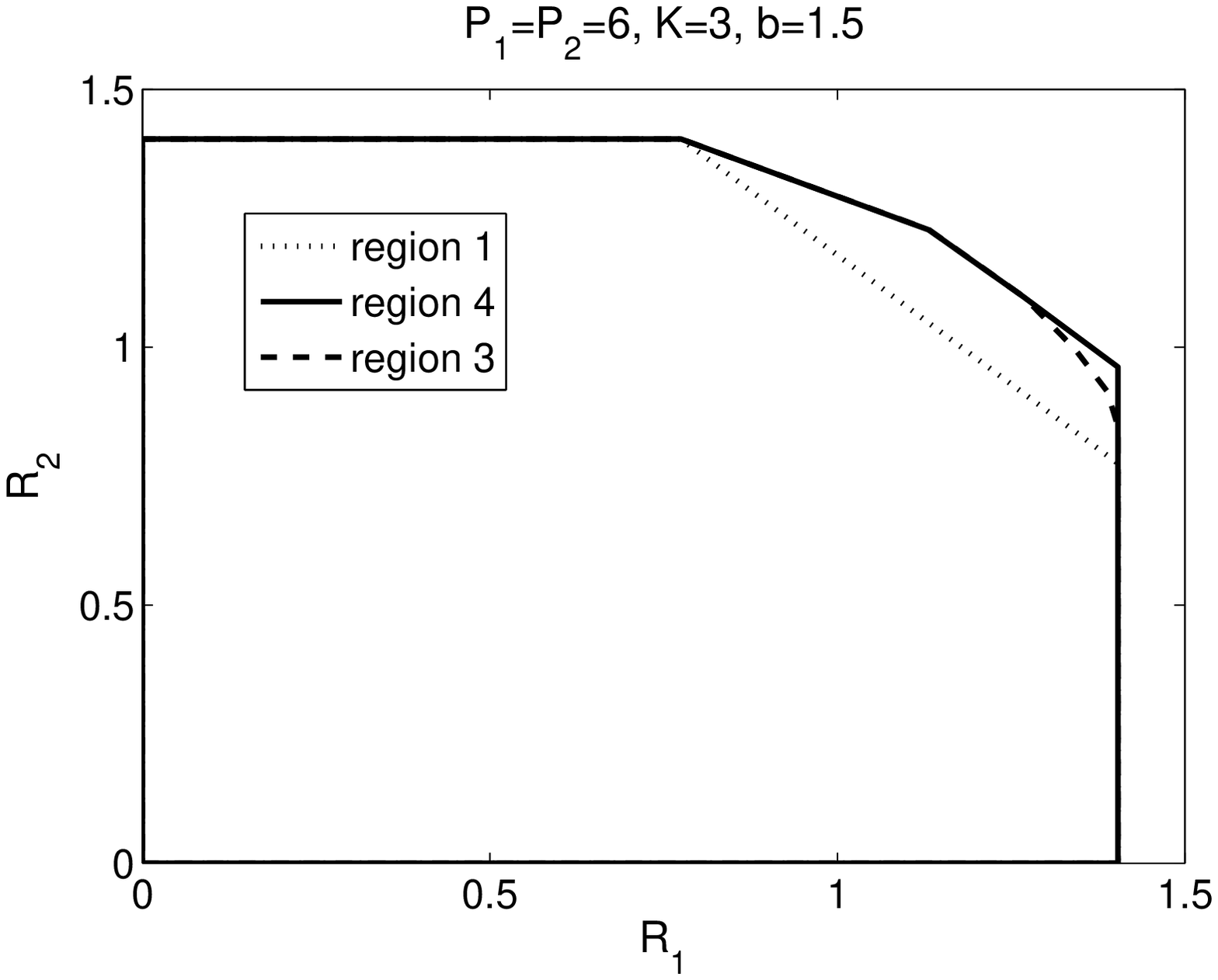}\\
(c)
\end{tabular}
\caption{\label{fig:4-6} Comparison of $\Rmat_1$, $\Rmat_3$ and
$\Rmat_4$. (a) $P_1=P_2=6, K=1.5, b=1.5$. (b) $P_1=P_2=6, K=2,
b=1.5$. (c) $P_1=P_2=6, K=3, b=1.5$.}
\end{figure}

The reason why $\Rmat_1$ can sometimes outperform $\Rmat_3$ is
that we apply sequential decoding at receiver 2, i.e., $Y_3$
decodes part of user 1's new message ($W_{12}$) first, then
decodes $W_2$. To improve the rate region, we can apply
simultaneous decoding for $W_{12}$ and $W_2$ as in the multiple
access channel, which leads to the following
constraints at receiver 2: \bqa R_{12}&\leq& I(W_{12};Y_3U_2)\label{eq:simul_1}\\
R_2&\leq& I(U_2;Y_3W_{12})-I(U_2;S_{11})\label{eq:simul_2}\\
R_{12}+R_2&\leq& I(U_2W_{12};Y_3)-I(U_2;S_{11})\label{eq:simul_3}
\eqa where $W_{12}$ is part of $W_1$ to be decoded by $Y_3$;
$S_{11}$ is the cell index of the message $W_{11}$, which is the
other part of $W_1$; $U_2$ is an auxiliary variable for dirty
paper coding $W_2$ against $S_{11}$. To evaluate
(\ref{eq:simul_1})-(\ref{eq:simul_3}), let $W_{12}\sim N(0,\beta
P_1)$, $S_{11}\sim N(0,(1-\alpha)P_1)$, $U_2=X_2+\mu S_{11}$,
where $X_2\sim N(0,P_2)$ and $\mu$ is a deterministic real number.
Thus, the right hand side of (\ref{eq:simul_1})-(\ref{eq:simul_3})
can be evaluated as \bqa
\!\!\!\!\!\!&\!\!\!\!\!\!&\!\!\!\!\!\!\zeta_1=\frac{1}{2}\log\left(\frac{(P_2+\mu^2\bar{\alpha}P_1)(P_2+b^2P_1+1)-(P_2+\mu
b\bar{\alpha}P_1)^2}{(P_2+\mu^2\bar{\alpha}P_1)(P_2+b^2\bar{\beta}P_1+1)-(P_2+\mu
b\bar{\alpha}P_1)^2}\right)\label{eq:simul4}\\
\!\!\!\!\!\!&\!\!\!\!\!\!&\!\!\!\!\!\!\zeta_2=\frac{1}{2}\log\left(\frac{P_2(P_2+b^2\bar{\beta}P_1+1)}{(P_2+\mu^2\bar{\alpha}P_1)(P_2+b^2\bar{\beta}P_1+1)-(P_2+\mu
b\bar{\alpha}P_1)^2}\right)\label{eq:simul5}\\
\!\!\!\!\!\!&\!\!\!\!\!\!&\!\!\!\!\!\!\zeta_3=\frac{1}{2}\log\left(\frac{P_2(P_2+b^2P_1+1)}{(P_2+\mu^2\bar{\alpha}P_1)(P_2+b^2\bar{\beta}P_1+1)-(P_2+\mu
b\bar{\alpha}P_1)^2}\right) \label{eq:simul6}
 \eqa Plugging (\ref{eq:simul4})-(\ref{eq:simul6}) into (\ref{eq:simul_1})-(\ref{eq:simul_3}),
 we can define a new region, $\Rmat_4$, based
on the idea of simultaneous decoding at receiver 2: \bqa R_1&\leq&
\gamma(K^2\alpha P_1)\\
R_1&\leq& \gamma(P_1)\\
R_1&\leq& \gamma(K^2(\alpha-\beta)P_1)+\gamma(\beta P_1)\\
R_1&\leq& \gamma(K^2(\alpha-\beta)P_1)+\zeta_1\\
R_1&\leq& \gamma((1-\beta)P_1)+\zeta_1\\
R_2&\leq& \zeta_2\\
R_1+R_2&\leq& \gamma(K^2(\alpha-\beta)P_1)+\zeta_3\\
R_1+R_2&\leq& \gamma((1-\beta)P_1)+\zeta_3
 \eqa
for all $0\leq\beta\leq\alpha\leq 1$, $-\infty<\mu<\infty$,
$\alpha+\bar{\alpha}=1$ and $\beta+\bar{\beta}=1$.
\begin{theorem} Region $\Rmat_4$ is achievable for the cognitive
ZIC.
\end{theorem}
Comparison of $\Rmat_1$, $\Rmat_3$ and $\Rmat_4$ is given in
Fig.\ref{fig:4-6}. It can be seen that for all values of $K$,
$\Rmat_4$ is always the superset of both $\Rmat_1$ and $\Rmat_3$.

%\begin{figure}[htp]
%\begin{tabular}{cc}
%\centerline{\leavevmode \epsfxsize=2.5 in \epsfysize=1.8 in
%\epsfbox{4.eps}} \\
%(a)\\
%\centerline{\leavevmode \epsfxsize=2.5 in \epsfysize=1.8 in
%\epsfbox{5.eps}}\\
%(b)\\
%\centerline{\leavevmode \epsfxsize=2.5 in \epsfysize=1.8 in
%\epsfbox{6.eps}}\\
%(c)
%\end{tabular} \caption{\label{fig:4-6} Comparison of $\Rmat_1$,
%$\Rmat_3$ and $\Rmat_4$. (a) $P_1=P_2=6, K=1.5, b=1.5$. (b)
%$P_1=P_2=6, K=2, b=1.5$. (c) $P_1=P_2=6, K=3, b=1.5$.}
%\end{figure}

It is worth noting that the two achievable regions $\Rmat_1$ and
$\Rmat_3$ have a common corner point \bqa
R_1=\gamma\left(\frac{b^2P_1}{1+P_2}\right),
R_2=C_2.\label{eq:corner point} \eqa It is conjectured that for
the classic ZIC, this is indeed the corner point of the capacity
region \cite{Costa:85IT,Sason:04IT}. It is not clear whether the
existence of the cognitive link may extend this corner point using
some other coding schemes for finite $K$.

\subsubsection{$b^2<1$} For the weak interference case, we only
know the sum rate capacity of the classic ZIC achieved at the
corner point \bqa R_1=C_1,
R_2=\gamma\left(\frac{P_2}{1+b^2P_1}\right).
 \eqa
The other corner point of the known achievable region for the ZIC
is described in (\ref{eq:corner point}).

\begin{figure}[htp]
\begin{tabular}{cc}
\leavevmode \epsfxsize=3.2in \epsfysize=2.4in \epsfbox{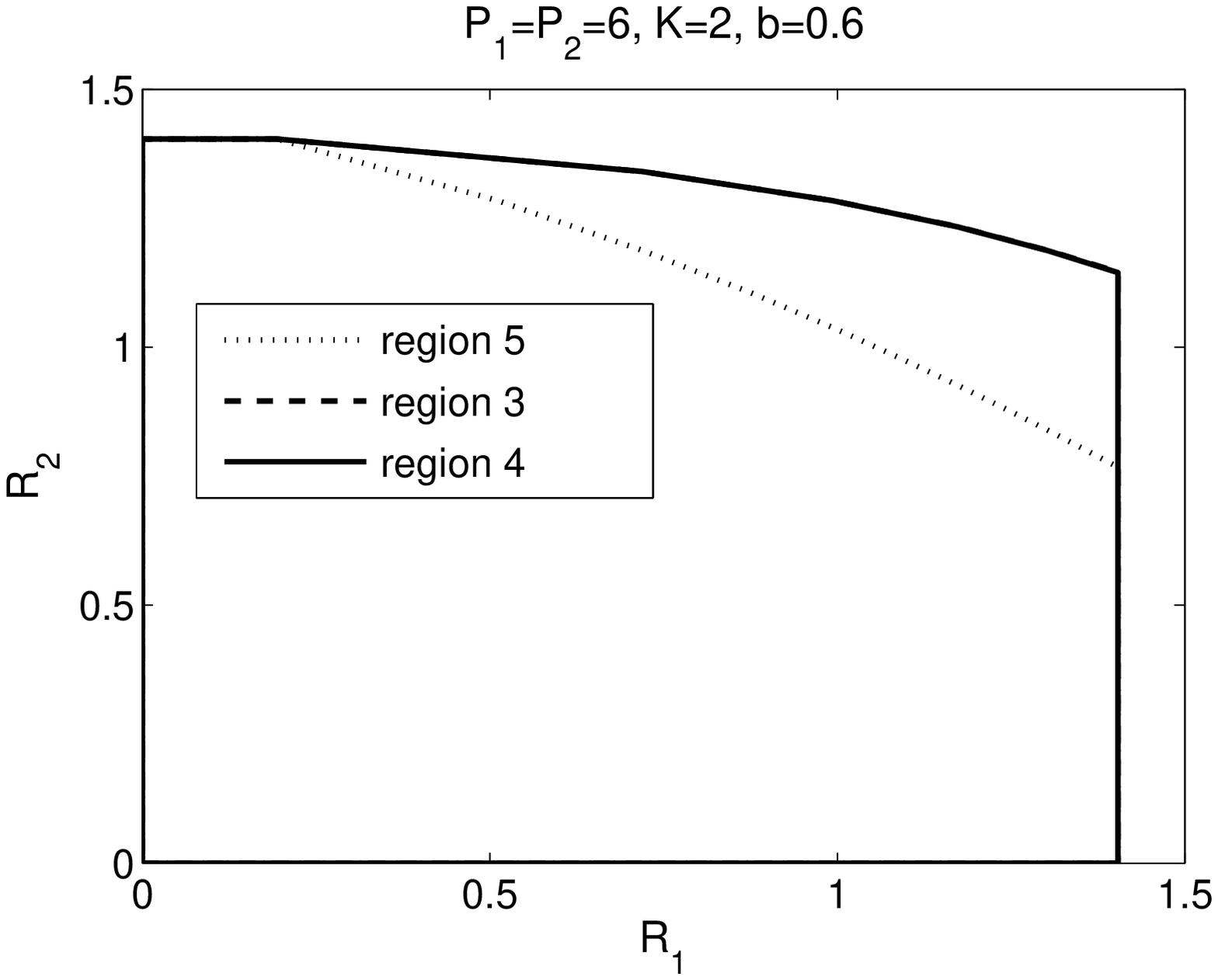} &
\leavevmode \epsfxsize=3.2in
\epsfysize=2.4in \epsfbox{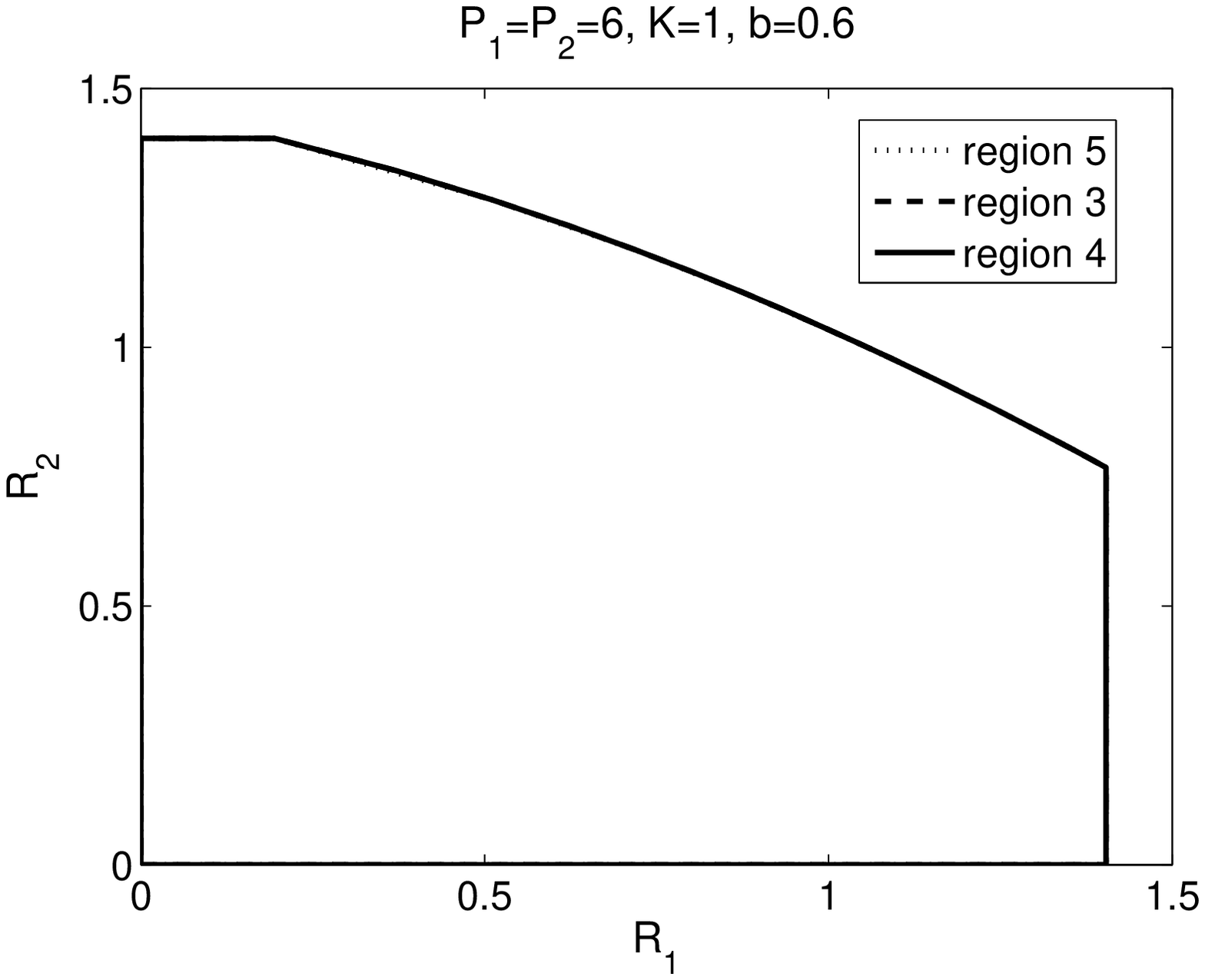}\\
(a) & (b)\\
\leavevmode \epsfxsize=3.2in \epsfysize=2.4in \epsfbox{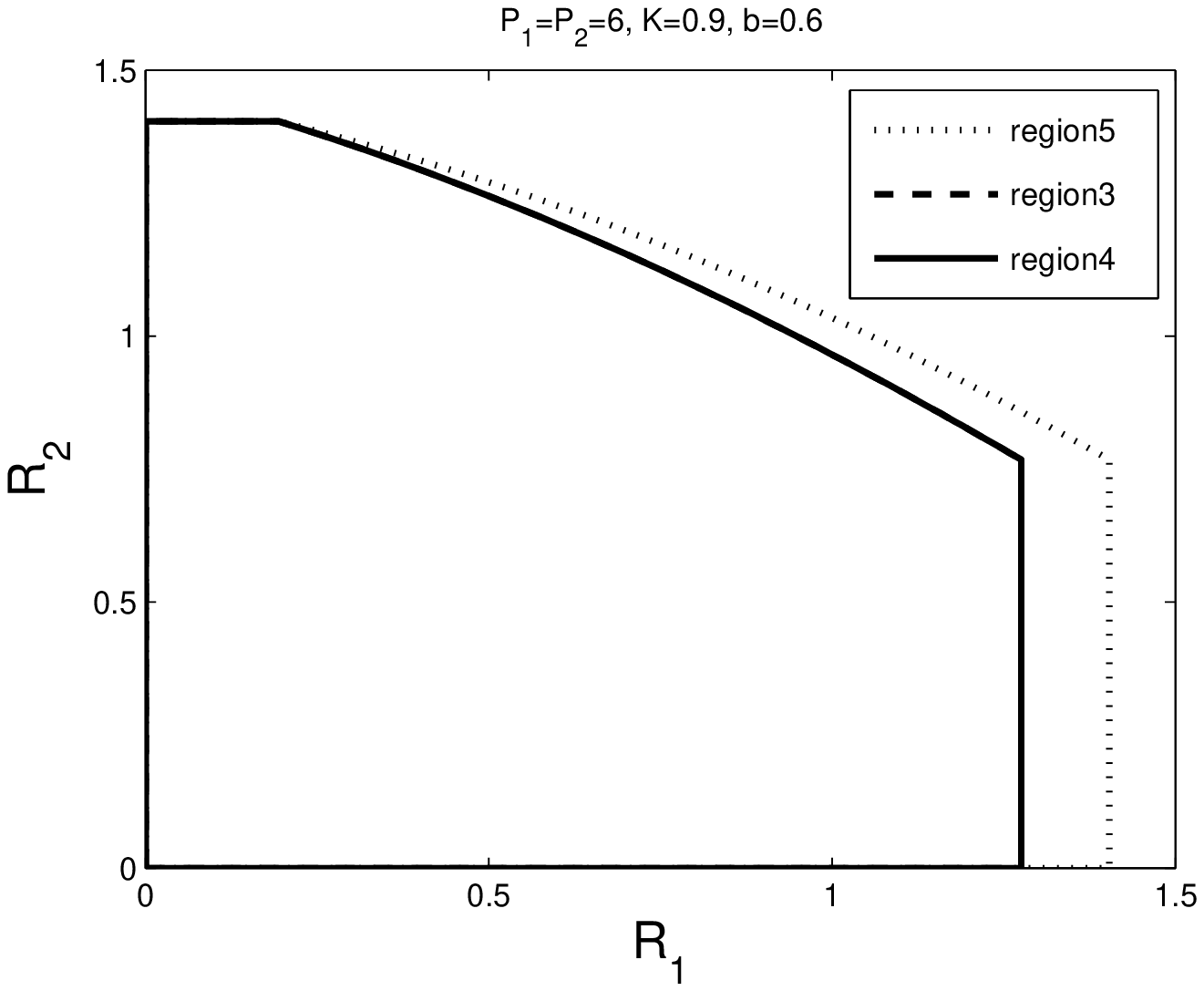}\\
(c)
\end{tabular}
\caption{\label{fig:7-9} Comparison of $\Rmat_3$, $\Rmat_5$ and
$\Rmat_5$. (a) $P_1=P_2=6, K=2, b=0.6$. (b) $P_1=P_2=6, K=1,
b=0.6$. (c) $P_1=P_2=6, K=0.9, b=0.6$.}
\end{figure}

Let us define $\Rmat_5$ as the Han-Kobayashi region
\cite{Han&Kobayashi:81IT} with $Q=\phi$ and Gaussian inputs for
the classical ZIC. Then, after Fourier-Motzkin elimination, and
removing redundant inequalities due to $b<1$, $\Rmat_5$ can be
expressed by \bqa R_1&\leq& \gamma(\alpha
P_1)+\gamma\left(\frac{b^2(1-\alpha)P_1}{1+b^2\alpha P_1+P_2}\right)\\
R_2&\leq& \gamma\left(\frac{P_2}{1+b^2\alpha P_1}\right) \eqa for
$\alpha\in [0,1]$. For the cognitive ZIC with weak interference,
the regions $\Rmat_3$ and $\Rmat_4$ are still achievable. We can
compare these three regions for different values of $K$, as
plotted in Fig.\ref{fig:7-9}.
%\begin{figure}[htp]
%\begin{tabular}{cc}
%\centerline{\leavevmode \epsfxsize=2.5 in \epsfysize=1.8 in
%\epsfbox{7.eps}} \\
%(a)\\
%\centerline{\leavevmode \epsfxsize=2.5 in \epsfysize=1.8 in
%\epsfbox{8.eps}}\\
%(b)\\
%\centerline{\leavevmode \epsfxsize=2.5 in \epsfysize=1.8 in
%\epsfbox{9.eps}}\\
%(c)
%\end{tabular} \caption{\label{fig:7-9} Comparison of $\Rmat_3$,
%$\Rmat_5$ and $\Rmat_5$. (a) $P_1=P_2=6, K=2, b=0.6$. (b)
%$P_1=P_2=6, K=1, b=0.6$. (c) $P_1=P_2=6, K=0.9, b=0.6$.}
%\end{figure}

When $K>1$, $\Rmat_3$ and $\Rmat_4$ are the same and they
outperform the HK region.

When $K=1$, regions $\Rmat_3$ and $\Rmat_4$ are indistinguishable
from $\Rmat_5$. In fact, we can prove that $\Rmat_3$ is indeed
inequivalent to $\Rmat_5$ for $b<1$ and $K=1$.

\begin{lemma}
$\Rmat_3$ is equivalent to $\Rmat_5$ for $b<1$ and $K=1$.
\end{lemma}
\begin{proof}
Since $K=1$, by removing redundant constraints, $\Rmat_3$ becomes
\bqa R_1\!\!&\leq&\!\!
\gamma(\alpha P_1)\label{eq:proof1_region3&5}\\
R_1\!\!&\leq&\!\!
\gamma((\alpha-\beta)P_1)+\gamma\left(\frac{b^2\beta
P_1}{1+b^2(1-\beta)P_1+P_2}\right)\label{eq:proof2_region3&5}\\
R_2\!\!&\leq&\!\!\gamma\left(\frac{P_2}{1+b^2(\alpha-\beta)P_1}\right)\label{eq:proof3_region3&5}
 \eqa
 We can easily verify that since $b<1$, for fixed $\alpha$, the right hand side of
 (\ref{eq:proof2_region3&5}) is a decreasing function of $\beta$.
Thus, when $\beta=0$, (\ref{eq:proof2_region3&5}) reaches its
 maximum value $\gamma(\alpha P_1)$. Therefore,
 (\ref{eq:proof1_region3&5}) is redundant and can be removed. Now,
 for the region defined by
 (\ref{eq:proof2_region3&5})-(\ref{eq:proof3_region3&5}), we can
 easily verify that for fixed $\beta$, the region increases with $\alpha$. Thus, the
 optimal operating point is for $\alpha=1$, i.e., transmitter 1
 uses all its power, which is intuitively true for
 $b<1$. By setting $\alpha=1$,
 (\ref{eq:proof2_region3&5})-(\ref{eq:proof3_region3&5}) reduces to
 $\Rmat_5$.
\end{proof}

When $K<1$, $\Rmat_3$ and $\Rmat_4$ are still the same, and they
are outperformed by the HK region. It is clear that, when $K\leq
1$, the idea of utilizing the cognitive link and applying dirty
paper coding to boost $R_2$ is strictly suboptimal for the
proposed coding scheme. Again, it is not clear if there is any
other coding scheme that can improve the rate region by utilizing
the cooperating link when $K\leq 1$.

\subsection{Capacity Outer Bounds} Besides the trivial outer bound
$\Rmat_2$ mentioned above, in this section, we propose a
nontrivial outer bound to the capacity region of the cognitive ZIC
for the case $b\leq 1$, and discuss some interesting implications
from this result. Before presenting the outer bound, let us first
introduce the following lemma.
\begin{lemma}\label{lemma:markov}
If $b\leq 1$, the capacity region of the cognitive ZIC
(\ref{eq:model_standard1})-(\ref{eq:model_standard3}) is the same
as
the cognitive ZIC given below: \bqa Y_1^{'}&=& X_1+Z_1\\
Y_2^{'}&=& KX_1+Z_2\\
Y_3^{'}&=& bY_1^{'}+X_2+\tilde{Z_3} \eqa where $\tilde{Z_3}\sim
N(0, 1-b^2)$ is independent of all the other random variables.
Thus, given $X_2$, \bqa X_1\Longrightarrow Y_1^{'}\Longrightarrow
Y_3^{'} \label{eq:Markov_ZIC} \eqa forms a Markov chain.
\end{lemma}
\begin{proof}
Since $Y_1$ does not cooperate with $Y_2$, $Y_3$, the capacity
region of the cognitive ZIC only depends on the marginal
distributions $p(y_1|x_1,x_2)$ and $p(y_2,y_3|x_1,x_2)$. Further,
since $Z_1$, $Z_2$ and $Z_3$ in the original model
(\ref{eq:model_standard1})-(\ref{eq:model_standard3}) are
independent, in the modified channel, $Z_2$ and $bZ_1+\tilde{Z_3}$
are also independent. Therefore, \bqa
p(y_2^{'},y_3^{'}|x_1,x_2)=p(y_2,y_3|x_1,x_2). \eqa Therefore, the
capacity region of the modified channel is the same as the
original channel, and the Markov chain (\ref{eq:Markov_ZIC})
follows automatically.
\end{proof}

With the lemma above, we are ready to derive our outer bound
described in the following theorem.
\begin{theorem}\label{thm:outer1}
Define $\Rmat_o$ to be the union of all nonnegative rate pairs
$(R_1, R_2)$ satisfying
\bqa R_1&\leq& I(X_1;Y_1Y_2|UX_2)\label{eq:outer1_eq1}\\
R_2&\leq& I(UX_2;Y_3)\label{eq:outer1_eq2} \eqa for all joint
distributions that factor as \bqa
p(u)p(x_1x_2|u)p(y_1y_2y_3|x_1x_2).\label{eq:U's Markov} \eqa Then
$\Rmat_o$ is a capacity outer bound for the cognitive ZIC.
\end{theorem}
\begin{proof}
Define $U_i=(Y_1^{i-1}, Y_2^{i-1}, W_2, X_2^{i-1})$, and we have
\bqa
nR_1&=& H(W_1|W_2)\\
&=&I(W_1;Y_1^n|W_2)+H(W_1|Y_1^n,W_2)\\
&\leq& I(W_1;Y_1^n|W_2)+n\epsilon_1 \label{eq:reason1}\\
&\leq& I(W_1;Y_1^nY_2^n|W_2)+n\epsilon_1\\
&=& \sum_{i=1}^n I(W_1;Y_{1i}Y_{2i}|Y_1^{i-1}Y_2^{i-1}W_2)+n\epsilon_1\\
&=& \sum_{i=1}^n
I(W_1;Y_{1i}Y_{2i}|Y_1^{i-1}Y_2^{i-1}W_2X_2^i)+n\epsilon_1 \label{eq:reason2}\\
&=&\sum_{i=1}^n I(X_{1i};Y_{1i}Y_{2i}|U_iX_{2i})+n\epsilon_1
\label{eq:reason3}
 \eqa
 (\ref{eq:reason1}) is due to Fano's inequality;
 (\ref{eq:reason2}) is because the codeword $X_{2i}$ is a function
 of $W_2$ and $Y_2^{i-1}$ as stated in (\ref{eq:encoding function
 2}); (\ref{eq:reason3}) is due to the memoryless channel model.
\bqa nR_2&=&H(W_2)=I(W_2;Y_3^n)+H(W_2|Y_3^n)\\
&\leq& I(W_2;Y_3^n)+n\epsilon_2\\
&=& \sum_{i=1}^n I(W_2;Y_{3i}|Y_3^{i-1})+n\epsilon_2\\
&=& \sum_{i=1}^n
\left\{h(Y_{3i}|Y_3^{i-1})-h(Y_{3i}|Y_3^{i-1}W_2)\right\}+n\epsilon_2\\
&\leq& \sum_{i=1}^n
\left\{h(Y_{3i})-h(Y_{3i}|Y_3^{i-1}W_2X_2^i)\right\}+n\epsilon_2\\
&\leq& \sum_{i=1}^n
\left\{h(Y_{3i})-h(Y_{3i}|Y_1^{i-1}W_2X_2^i)\right\}+n\epsilon_2\label{eq:reason4}\\
&\leq& \sum_{i=1}^n
\left\{h(Y_{3i})-h(Y_{3i}|U_iX_{2i})\right\}+n\epsilon_2\\
&=& \sum_{i=1}^n I(U_iX_{2i};Y_{3i})+n\epsilon_2 \eqa
(\ref{eq:reason4}) follows from Lemma \ref{lemma:markov}. When
$b\leq 1$, the capacity region of the cognitive ZIC is equivalent
to the cognitive ZIC such that given $X_2$, \bqa
X_1\Longrightarrow Y_1\Longrightarrow Y_3.
\label{eq:Markov2_ZIC}\eqa Therefore it suffices to establish the
outer bound for the cognitive ZIC satisfying
(\ref{eq:Markov2_ZIC}). Due to (\ref{eq:Markov2_ZIC}),
conditioning on $(X_2^{i-1},Y_1^{i-1})$, the random vector
$Y_3^{i-1}$ is independent of all other variables, including
$Y_{3i}$. Thus, given $X_2^{i}, W_2$, \bqa Y_{3i}\Longrightarrow
Y_1^{i-1}\Longrightarrow Y_3^{i-1}. \eqa Hence, \bqa
h(Y_{3i}|Y_3^{i-1}W_2X_2^i)\geq h(Y_{3i}|Y_1^{i-1}W_2X_2^i). \eqa
Thus, (\ref{eq:reason4}) follows.

Now we introduce the time sharing random variable $I$ to be
independent of all other variables and uniformly distributed over
$\{1, 2, \cdot\cdot\cdot, n\}$. Define $U=(U_I,I)$, $X_1=X_{1I}$,
$X_2=X_{2I}$, $Y_1=Y_{1I}$, $Y_2=Y_{2I}$ and $Y_3=Y_{3I}$. Then,
we have \bqa nR_1&\leq& \sum_{i=1}^n
I(X_{1i};Y_{1i}Y_{2i}|U_iX_{2i})+n\epsilon_1\\
&=& n I(X_{1I};Y_{1I}Y_{2I}|U_IX_{2I}I)+n\epsilon_1\\
&=& n I(X_1;Y_1Y_2|UX_2)+n\epsilon_1\\
nR_2&\leq&\sum_{i=1}^n I(U_iX_{2i};Y_{3i})+n\epsilon_2\\
&=& n I(U_IX_{2I};Y_{3I}|I)+n\epsilon_2\\
&\leq& n I(IU_IX_{2I};Y_{3I})+n\epsilon_2\\
&=& nI(UX_2;Y_3)+n\epsilon_2
 \eqa
 This establishes (\ref{eq:outer1_eq1})-(\ref{eq:outer1_eq2}) but it requires $Y_3$
satisfy Markov chain (\ref{eq:Markov2_ZIC}), i.e., the noise
$Z_3=bZ_1+\tilde{Z}_3$ as in Lemma \ref{lemma:markov}. However,
due to the special form of
(\ref{eq:outer1_eq1})-(\ref{eq:outer1_eq2}), we now show that this
outer bound can be equivalently evaluated by assuming $Z_3$ to be
independent of $Z_1$ as in the original channel model. According
to the channel model and the Markov chain (\ref{eq:U's Markov}),
$(U, X_1, X_2)$ are independent of all the noises $Z_1, Z_2$ and
$Z_3$. Therefore, to compute (\ref{eq:outer1_eq2}), \bqa
I(UX_2;Y_3)&=&h(bX_1+X_2+Z_3)-h(bX_1+Z_3|UX_2)\\
&=&h(bX_1+X_2+Z_3)-h(b(X_1|UX_2)+Z_3) \eqa Thus, any value that
$I(UX_2;Y_3)$ can achieve with $Z_3=bZ_1+\tilde{Z}_3$ can also be
achieved with $Z_3$ independent of $Z_1$. Therefore, The
relaxation of $Z_3$ to its original model will yield the same
outer bound. This proves Theorem \ref{thm:outer1}.
\end{proof}

To recap, the proof of Theorem \ref{thm:outer1} is accomplished in
two steps. The first step utilizes Lemma \ref{lemma:markov} and
establishes the outer bound for the equivalent channel satisfying
the Markov chain conditions \ref{eq:Markov2_ZIC}. In the second
step, we prove that the the outer bound can be equivalently
evaluated using the original channel model. Theorem
\ref{thm:outer1} is valid for all cognitive ZIC as long as $b\leq
1$. This outer bound is analogous to the capacity region of a
degraded broadcast channel where receiver $Y_3$ sees a degraded
channel compared with that of $(Y_1, Y_2)$.

Next, we derive another outer bound for the specific case where
$b\leq 1$ and $K\geq 1$. Before that, we introduce another lemma.
\begin{lemma}\label{lemma:modify Y_1}
If $K\geq 1$, the capacity region of the cognitive ZIC
(\ref{eq:model_standard1})-(\ref{eq:model_standard3}) is the same
as
the cognitive ZIC given below: \bqa \tilde{Y_1}&=& \frac{1}{K}\tilde{Y_2}+\tilde{Z_1}\\
\tilde{Y_2}&=& KX_1+Z_2\\
\tilde{Y_3}&=& bX_1+X_2+Z_3\eqa where $\tilde{Z_1}\sim N(0,
1-\frac{1}{K^2})$ is independent of all the other random
variables. Thus, \bqa X_1\Longrightarrow
\tilde{Y_2}\Longrightarrow \tilde{Y_1}. \label{eq:Markov3_ZIC}
\eqa Further, for all random variable $U$ such that
$U\Longrightarrow (X_1,X_2)\Longrightarrow
(\tilde{Y_1},\tilde{Y_2},\tilde{Y_3})$, \bqa X_1\Longrightarrow
(U,X_2,\tilde{Y_2})\Longrightarrow \tilde{Y_1}.
\label{eq:Markov4_ZIC}\eqa
\end{lemma}
\begin{proof}
The argument is similar to that of Lemma \ref{lemma:markov}. Since
$Y_1$ does not cooperate with $Y_2$ or $Y_3$, the capacity region
only depends on $p(y_1|x_1x_2)$ and $p(y_2,y_3|x_1,x_2)$.
Therefore, making the noise at $Y_1$ to be correlated with that of
$Y_2$ will have no effect on the capacity region. Thus,
(\ref{eq:Markov3_ZIC}) holds. Furthermore, since $\tilde{Z_1}$ is
independent of all other variables, (\ref{eq:Markov4_ZIC}) also
holds.
\end{proof}

When $K\geq 1$, according to Lemma \ref{lemma:modify Y_1}, we only
need to consider the cognitive ZIC such that
(\ref{eq:Markov4_ZIC}) is satisfied. Since $b\leq 1$ and due to
the fact that Theorem \ref{thm:outer1} holds for the original
cognitive ZIC channel model, the outer bound
(\ref{eq:outer1_eq1})-(\ref{eq:outer1_eq2}) is still an outer
bound for the system defined in Lemma \ref{lemma:modify Y_1}. For
all $U$ under condition (\ref{eq:U's Markov}), due to
(\ref{eq:Markov4_ZIC}), \bqa I(X_1;Y_1|UX_2Y_2)=0. \eqa Thus, we
can rewrite the outer bound $\Rmat_o$ as \bqa R_1&\leq&
I(X_1;Y_2|UX_2)\label{eq:outer3}\\
R_2&\leq& I(UX_2;Y_3)\label{eq:outer4}. \eqa

Next, we give the second outer bound in the theorem below.
\begin{theorem}\label{thm:outer2}
If $b\leq 1$ and $K\geq 1$, an outer bound to the capacity region
of the cognitive ZIC is the union of all nonnegative rate pairs
$(R_1, R_2)$
satisfying \bqa R_1&\leq& C_1\label{eq:outer_1}\\
R_2&\leq& C_2\label{eq:outer_2}\\
R_1&\leq& \gamma(K^2\alpha P_1)\label{eq:outer_3}\\
R_2&\leq&
\gamma\left(\frac{b^2\bar{\alpha}P_1+P_2+2b\sqrt{\bar{\alpha}P_1P_2}}{1+b^2\alpha
P_1}\right)\label{eq:outer_4} \eqa for $0\leq \alpha\leq 1$ and
$\alpha+\bar{\alpha}=1$.
\end{theorem}
\begin{proof}
(\ref{eq:outer_1}) and $(\ref{eq:outer_2})$ are trivial outer
bounds. (\ref{eq:outer_3}) and (\ref{eq:outer_4}) are derived from
(\ref{eq:outer3}) and (\ref{eq:outer4}), respectively. Consider
(\ref{eq:outer3}).
\bqa h(Y_2|UX_2)&=&h(KX_1+Z_2|UX_2)\\
&\leq& h(KX_1+Z_2)\\
&\leq& \frac{1}{2}\log(2\pi e(1+K^2P_1)) \eqa On the other hand,
\bqa
h(Y_2|UX_2)&\geq& h(Y_2|UX_1X_2)\\
&=&h(Z_2)=\frac{1}{2}\log(2\pi e) \eqa Without loss of generality,
we set \bqa h(Y_2|UX_2)=\frac{1}{2}\log(2\pi e(1+K^2\alpha
P_1))\label{eq:outerproof1} \eqa where $\alpha\in [0,1]$. Thus,
bound (\ref{eq:outer3}) becomes \bqa R_1&\leq&\frac{1}{2}\log(2\pi
e(1+K^2\alpha P_1)). \eqa By Lemma 1 of \cite{Thomas:87IT}, \bqa
h(Y_2|UX_2)&\leq&
h(KX_1+Z_2|X_2)\\
&\leq& h(KX_1^*+Z_2|X_2^*)\\
&=&\frac{1}{2}\log(2\pi
e(1+K^2\mbox{Var}(X_1^*|X_2^*)))\label{eq:outerproof2} \eqa where
$X_1^*$ and $X_2^*$ are Gaussian distributed variables with the
same covariance matrix with that of $X_1$ and $X_2$. Combining
(\ref{eq:outerproof1}) and (\ref{eq:outerproof2}), we obtain \bqa
\mbox{Var}(X_1^*|X_2^*)\geq \alpha P_1 \label{eq:outerproof3}\eqa
Since, \bqa \mbox{Var}(X_1^*|X_2^*)=
E[(X_1^*)^2]-E[(E[X_1^*|X_2^*])^2],
 \eqa Combined with (\ref{eq:outerproof3}), we obtain \bqa
E[(E[X_1^*|X_2^*])^2]&\leq& \bar{\alpha}P_1. \eqa Thus,
\bqa E(X_1X_2)&=&E(X_1^*X_2^*)\\
&\leq&
\left(E[(E[X_1^*|X_2^*])^2]E[(X_2^*)^2]\right)^{\frac{1}{2}}\\
&\leq&\sqrt{\bar{\alpha}P_1P_2}. \eqa Next we consider
(\ref{eq:outer4}). Since $K\geq 1$ and $b\leq 1$, given $X_2$,
$Y_3$ is a degraded version of $Y_2$. By the entropy power
inequality, \bqa 2^{2h(Y_3|UX_2)}\!\!\!\!\!&\geq&\!\!\!\!\!
2^{2h(\frac{b}{K}Y_2|UX_2)}+2^{2h(Z^{'})}\\
\!\!\!\!\!&=&\!\!\!\!\!\frac{b^2}{K^2}2^{2h(Y_2|UX_2)}+2\pi e(1-\frac{b^2}{K^2})\\
\!\!\!\!\!&=&\!\!\!\!\!\frac{b^2}{K^2}2\pi e(1+K^2\alpha P_1)+2\pi
e(1-\frac{b^2}{K^2})\\
\!\!\!\!\!&=&\!\!\!\!\! 2\pi e(1+b^2\alpha P_1) \eqa Therefore,
\bqa h(Y_3|UX_2)\geq \frac{1}{2}\log(2\pi e(1+b^2\alpha P_1)) \eqa
Thus, (\ref{eq:outer4})
becomes \bqa I(UX_2;Y_3)\!\!\!\!\!&=&\!\!\!\!\!h(Y_3)-h(Y_3|UX_2)\\
\!\!\!\!\!&\leq&\!\!\!\!\!
\frac{1}{2}\log\left(\frac{b^2P_1+P_2+2b\sqrt{\bar{\alpha}P_1P_2}+1}{1+b^2\alpha
P_1}\right)  \eqa This completes the proof of Theorem
\ref{thm:outer2}.
\end{proof}

\begin{corollary}
If $K\leq 1$ and $b\leq 1$, then
$(C_1,\gamma(\frac{P_2}{1+b^2P_1}))$ is the corner point of the
capacity region for the cognitive Z channel.
\end{corollary}
\begin{proof}
The achievability of this point is trivial, as this is indeed the
sum-rate capacity of the classical ZIC, i.e., one can achieve this
point in the absence of the cooperating link.

For the converse part, if $b\leq 1$, when $K=1$ and $R_1=C_1$,
according to Theorem \ref{thm:outer2},
$R_2\leq\gamma(\frac{P_2}{1+b^2P_1})$. Thus, this corner point is
on the boundary of the capacity region of the cognitive ZIC when
$K=1$. Since the capacity outer bound for the case with $K=1$ is
also an outer bound for the case with $K<1$ if all other
parameters remain the same, this is also a corner point of the
capacity region for all $K\leq 1$.
\end{proof}

\begin{figure}[htp]
\begin{tabular}{cc}
\leavevmode \epsfxsize=3.2in \epsfysize=2.4in \epsfbox{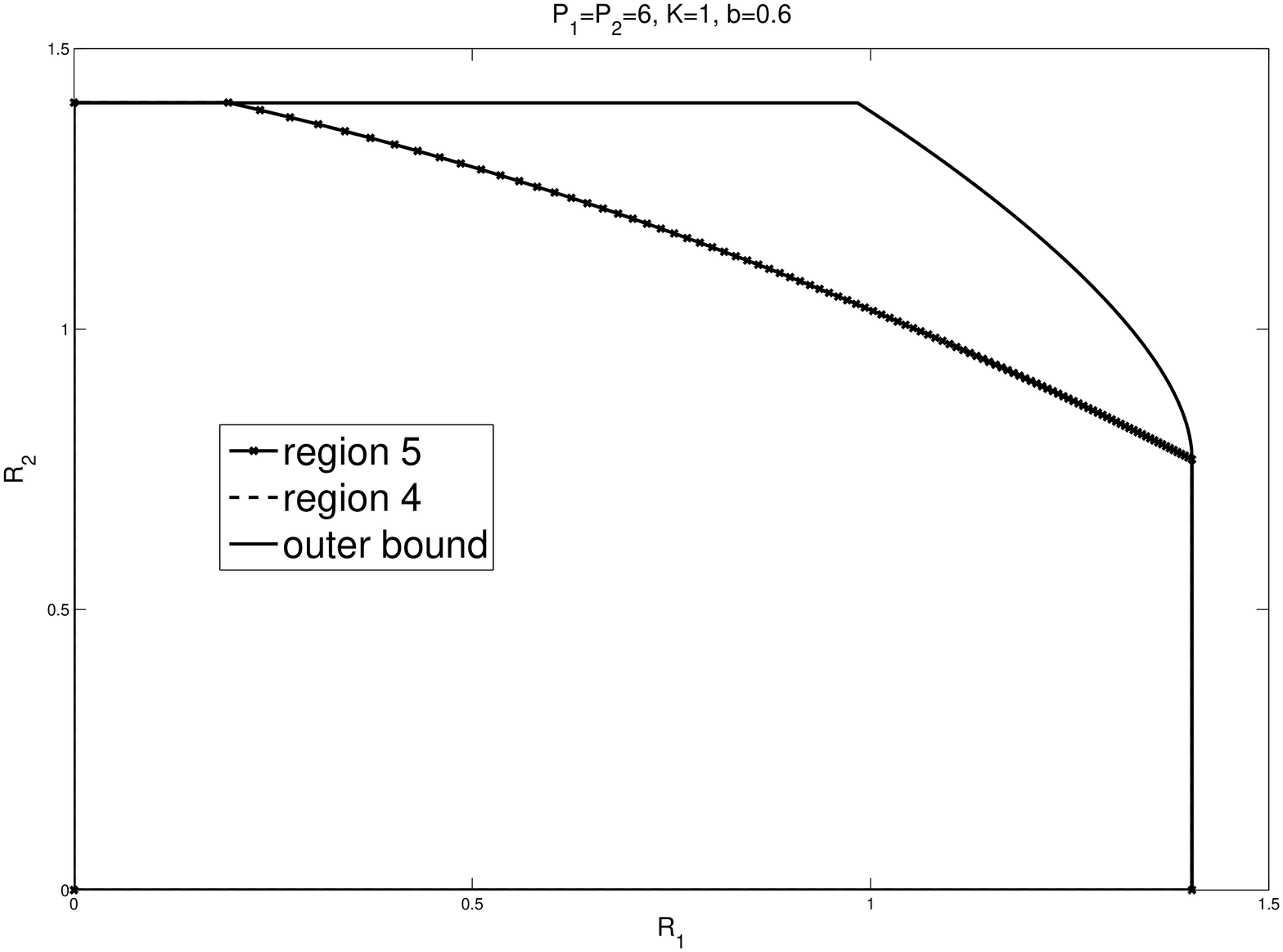}
& \leavevmode \epsfxsize=3.2in
\epsfysize=2.4in \epsfbox{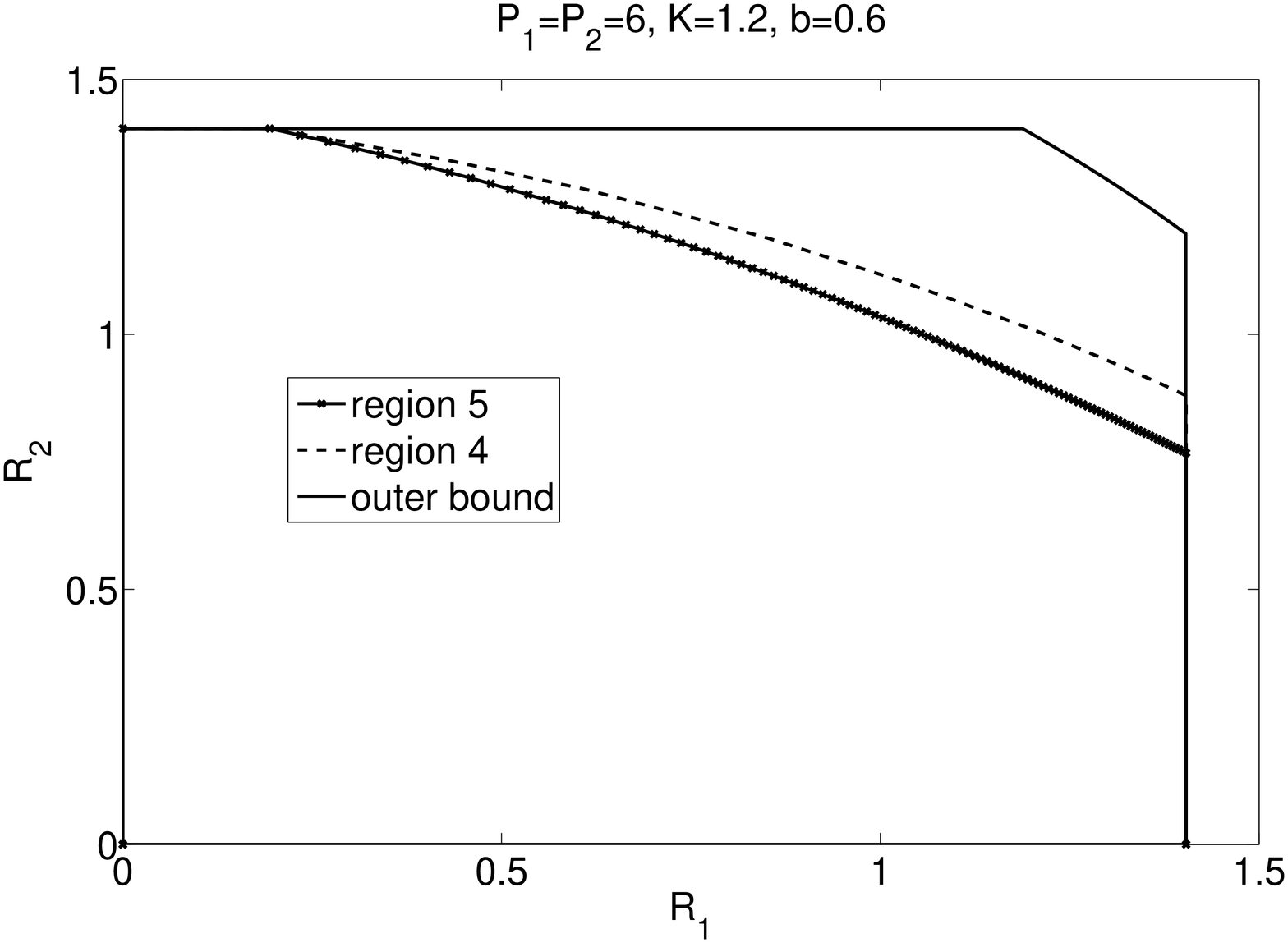}\\
(a) & (b)
\end{tabular}
\caption{\label{fig:outer} Comparison of $\Rmat_5$, $\Rmat_4$ and
$\Rmat_o$. (a) $P_1=P_2=6, K=1, b=0.6$. (b) $P_1=P_2=6, K=1.2,
b=0.6$.}
\end{figure}

That is to say, when the interference is weak ($b\leq 1$) and the
cooperating link is weak ($K\leq 1$), the cooperating link becomes
useless when user 1 is transmitting at its maximum rate. While
this does not imply that the entire capacity region will not be
affected by the cooperating link, this corner point is
particularly important in practice: it is often desirable in a
cognitive radio system with primary-secondary user pairs that the
primary user's rate is not affected by the interference of the
secondary user. We note that in the absence of the cooperating
link, this is also the sum-rate capacity corner point. However, no
such statement can be made for the cognitive Z channel as the
slope of the outer bound at this corner point is not necessarily
smaller than $45^o$. The comparison of the outer bound $\Rmat_o$
and the inner bounds are given in Fig. \ref{fig:outer}. Note that
$\Rmat_4$ and $\Rmat_5$ coincide with each other in (a) when $K=1$
while $\Rmat_4$ outperforms $\Rmat_5$ in (b) when $K>1$. Also,
when $K=1$, the optimal corner point
$(C_1,\gamma(\frac{P_2}{1+b^2P_1}))$ is shown in (a).

\section{Conclusion and Discussion}
This paper studied the interference channel with one cognitive
transmitter (ICOCT), from both the noncausal and causal
perspectives. For the noncausal ICOCT (or interference channel
with degraded message sets), we proposed a new achievable rate
region which generalizes existing results reported in the
literature. The proposed coding scheme leads to a rate region that
reduces to Marton's achievable rate region for the general
broadcast channels in the absence of the primary transmitter. This
proposed achievable region, together with a new outer bound,
establishes the capacity region of a class of semi-deterministic
IC-DMS.

%For the causal ICOCT, we derived the inner and outer bounds for
%both the strong and weak interference cases. We also extended the
%proposed bounds to the Gaussian channels and evaluated them in
%numerical examples.
%
% For the cognitive ZIC, without any degradedness assumptions, we
%proposed nontrivial inner and outer bounds for the capacity
%region, which lead to the establishment of a corner point of the
%capacity region when $K=1$ and $b<1$. On the other hand, in the
%achievable region, the cognitive ZIC still has the same corner
%point as Costa's corner point for the classic ZIC (when
%$R_2=C_2$). However, we still cannot prove if this is the corner
%point of the capacity region.

The causal ICOCT, which is a special case of interference channels
with generalized feedback, imposes causality constraint on the way
the cognitive transmitter can cooperate with the primary user.
Motivated by practical constraint on the so-called interference
temperature in the spectrum sharing cognitive radio system, we
focus on the Gaussin Z interference channel (ZIC) in which the
interference link from the cognitive transmitter and the primary
receiver is negligible. Both capacity inner bounds and outer
bounds are proposed for the cognitive Gaussian ZIC. The capacity
bounds established some intuitive results with regard to the
usefulness of the cognitive capability. In general, when the
feedback link is weak, it is generally not useful to utilize the
cognitive capability, as far as the capacity region is concerned.

\begin{appendix}

%\section{Proof of theorem 1}
{\em A. Proof of Theorem \ref{thm:new inner bound}}

\begin{proof}
{\bf Codebook generation:} Generate $2^{nR_{10}}$ independent and
identically distributed (i.i.d.) codewords $\ubf_{10}(j_{10})$,
$j_{10}=1,\cdots,2^{nR_{10}}$, according to
$\prod_{i=1}^{n}p(u_{10,i})$. For each codeword
$\ubf_{10}(j_{10})$, generate $2^{nR_{11}}$ i.i.d. codewords
$\ubf_{11}(j_{11},j_{10})$, $j_{11}=1,\cdots,2^{nR_{11}}$,
according to $\prod_{i=1}^n p(u_{11,i}|u_{10,i})$. For each pair
of $(\ubf_{11}, \ubf_{10})$, generate one codeword
$\xbf_1(j_{11},j_{10})$ according to $\prod_{i=1}^n
p(x_{1,i}|u_{11,i},u_{10,i})$.

For each codeword $\ubf_{10}(j_{10})$, also generate $2^{nL_{20}}$
i.i.d. codewords $\vbf_{20}(l_{20},j_{10})$,
$l_{20}=1,\cdots,2^{nL_{20}}$, according to $\prod_{i=1}^n
p(v_{20,i}|u_{10,i})$ and randomly place them into $2^{nR_{20}}$
bins. For each codeword pair $(\vbf_{20}(l_{20},j_{10}),
\ubf_{10}(j_{10}))$, generate $2^{nL_{11}}$ i.i.d. codewords
$\vbf_{11}(l_{11},l_{20},j_{10})$, $l_{11}=1,\cdots,2^{nL_{11}}$,
according to $\prod_{i=1}^n p(v_{11,i}|v_{20,i},u_{10,i})$ and
randomly place them into $2^{nR_{11}}$ bins. For each codeword
pair $(\vbf_{20}(l_{20},j_{10}),\ubf_{10}(j_{10}))$, generate
$2^{nL_{22}}$ i.i.d. codewords $\vbf_{22}(l_{22},l_{20},j_{10})$,
$l_{22}=1,\cdots,2^{nL_{22}}$, according to $\prod_{i=1}^n
p(v_{22,i}|v_{20,i},u_{10,i})$ and randomly place them into
$2^{nR_{22}}$ bins. For each set
$(\vbf_{22},\vbf_{20},\vbf_{11},\ubf_{11},\ubf_{10})$, generate
one codeword $\xbf_2(l_{22},l_{20},l_{11},j_{11},j_{10})$
according to \\
$\prod_{i=1}^n
p(x_{2,i}|v_{22,i},v_{20,i},v_{11,i},u_{11,i},u_{10,i})$.

{\bf Encoding:} For user 1 to send message $(j_{11},j_{10})$, it
simply transmits the codeword $\xbf_1(j_{11},j_{10})$.

Suppose user 2 wants to send message $(j_{22},j_{20})$, encoder 2
first looks into bin $j_{20}$ for codeword
$\vbf_{20}(l_{20},j_{10})$ such that \bqa
(\vbf_{20}(l_{20},j_{10}), \ubf_{11}(j_{11},j_{10}),
\ubf_{10}(j_{10}))\in T_{\epsilon}(V_{20}, U_{11},
U_{10})\label{event:V20, U11 joint typical} \eqa where
$T_{\epsilon}(\cdot)$ denotes jointly typical set. After it finds
$\vbf_{20}(l_{20},j_{10})$, it looks for a codeword
$\vbf_{22}(l_{22},l_{20},j_{10})$ in bin $j_{22}$ and a codeword
$\vbf_{11}(l_{11},l_{20},j_{10})$ in bin $j_{11}$ such that \bqa
(\vbf_{22}(l_{22},l_{20},j_{10}),\vbf_{11}(l_{11},l_{20},j_{10}),\ubf_{11}(j_{11},j_{10}),\vbf_{20}(l_{20},j_{10}),\ubf_{10}(j_{10}))\in
T_{\epsilon}(V_{22}V_{11}U_{11}V_{20}U_{10}) \label{event:V22 V11
joint typical}\eqa If there are more than one such codewords, pick
the one with the smallest index; if there is no such codeword,
declare an error. Then, user 2 sends
$\xbf_2(l_{22},l_{20},l_{11},j_{11},j_{10})$. The diagram of the
encoding scheme is illustrated in Fig. \ref{fig:encoding}.

{\bf Decoding:} Given $\ybf_1$, receiver 1 looks for all sequences
$\vbf_{11}$, $\ubf_{11}$, $\ubf_{10}$, $\vbf_{20}$ such that \bqa
(\vbf_{11}, \ubf_{11}, \ubf_{10}, \vbf_{20}, \ybf_1)\in
T_{\epsilon}(V_{11}U_{11}U_{10}V_{20}Y_1)\label{event:joint
typical at dec1} \eqa If all the codewords $\vbf_{11}$ have the
same bin indices $\hat{j}_{11}$, all $\ubf_{11}$ and $\ubf_{10}$
have the same message indices $\hat{\hat{j}}_{11}$ and
$\hat{j}_{10}$ respectively, and if
$\hat{j}_{11}=\hat{\hat{j}}_{11}$, we declare
$j_{11}=\hat{j}_{11}$, $j_{10}=\hat{j}_{10}$; otherwise, declare
an error.

Given $\ybf_2$, receiver 2 looks for all sequences $\vbf_{22}$,
$\vbf_{20}$, $\ubf_{10}$ such that \bqa (\vbf_{22}, \vbf_{20},
\ubf_{10}, \ybf_2)\in
T_{\epsilon}(V_{22}V_{20}U_{10}Y_2)\label{event:joint typical at
dec2} \eqa If all the codewords $\vbf_{22}$ and $\vbf_{20}$ have
the same bin indices $\hat{j}_{22}$ and $\hat{\hat{j}}_{20}$
respectively, we declare $j_{22}=\hat{j}_{22}$,
$j_{20}=\hat{\hat{j}}_{20}$; otherwise, declare an error.

{\bf Error analysis:} By the symmetry of random code generation,
without loss of generality, we assume the messages $(j_{11},
j_{10}, j_{22}, j_{20})=(1, 1, 1, 1)$ are sent. Let $P_{e,enc2}$,
$P_{e,dec1}$, $P_{e,dec2}$ denote the error probabilities at
encoder 2, decoder 1 and decoder 2 respectively. Denote the
codeword $\vbf_{20}(l_{20},j_{10})$ as
$\vbf_{20}(j_{20},k_{20},j_{10})$, where $j_{20}$ is its bin index
and $k_{20}$ is its index within the bin. Similarly, we denote
$\vbf_{11}(l_{11},l_{20},j_{10})$ and
$\vbf_{22}(l_{22},l_{20},j_{10})$ as
$\vbf_{11}(j_{11},k_{11},j_{20},k_{20},j_{10})$ and
$\vbf_{22}(j_{22},k_{22},j_{20},k_{20},j_{10})$ respectively.

1) Error occurs at encoder 2 when one or both of the following
events occur. \beq\begin{array}{ll} E_1: \mbox{there is no
$\vbf_{20}$ such that (\ref{event:V20, U11 joint
typical}) holds.}\\
E_2: \mbox{there is no pair $(\vbf_{22},\vbf_{11})$ such that
(\ref{event:V22 V11 joint typical}) holds.}
\end{array}\eeq
where event $E_1$ can be further divided into two sub-events:
\beq\begin{array}{ll} E_{11}: \mbox{$\ubf_{11}(j_{11},j_{10})$ and $\ubf_{10}(j_{10})$ are not jointly typical.}\\
E_{12}: \mbox{With $\ubf_{11}(j_{11},j_{10})$ and
$\ubf_{10}(j_{10})$ jointly typical, there is no $\vbf_{20}$ that
satisfies (\ref{event:V20, U11 joint typical}).}
\end{array}\eeq
According to the code book generation, it is obvious that $
P(E_{11})\leq \epsilon.$ For $P(E_{12})$, we have
 \bqa
P(E_{12})&=&(1-P[(\vbf_{20}(1,k_{20},1),\ubf_{11}(1,1),\ubf_{10}(1))\in
T_{\epsilon}(V_{20}U_{11}U_{10})])^{2^{n(L_{20}-R_{20})}}\\
&\leq&
(1-2^{-n(I(V_{20};U_{11}|U_{10})+\epsilon)})^{2^{n(L_{20}-R_{20})}}\\
&\leq&
\exp(-2^{n(L_{20}-R_{20}-I(V_{20};U_{11}|U_{10})-\epsilon)}) \eqa
So, (\ref{eq:rate 1}) guarantees $P(E_1)\rightarrow 0$ as
$n\rightarrow \infty$.

Event $E_2$ can be divided into the following three sub-events:
\beq\begin{array}{ll} E_{21}: \mbox{There is no codeword
$\vbf_{11}$ such that
$(\vbf_{11},\ubf_{11},\vbf_{20},\ubf_{10})\in
T_{\epsilon}(V_{11}U_{11}V_{20}U_{10})$.}\\
E_{22}: \mbox{There is no codeword $\vbf_{22}$ such that
$(\vbf_{22},\ubf_{11},\vbf_{20},\ubf_{10})\in
T_{\epsilon}(V_{22}U_{11}V_{20}U_{10})$.}\\
E_{23}: \mbox{Provided $E_{21}^c$ and $E_{22}^c$ occur, there is
no pair $(\vbf_{22}, \vbf_{11})$ such that (\ref{event:V22 V11
joint typical}) holds.}
\end{array}\eeq
Following similar derivations as for event $E_{12}$, the error
probability for events $E_{21}$ and $E_{22}$ will be arbitrarily
small as $n\rightarrow \infty$ if (\ref{eq:rate 2.1}) and
(\ref{eq:rate 2.2}) hold respectively. By using the second moment
method as in \cite{ElGamal&VanderMeulen:81IT}, one can also show
that $P(E_{23})\rightarrow 0$ as $n\rightarrow \infty$ if
(\ref{eq:rate 2}) holds. Since \bqa P_{e,enc2}=P\{E_1\cup
E_2\}\leq P(E_1)+P(E_2) \eqa (\ref{eq:rate 1}) and (\ref{eq:rate
2}) ensure $P_{e,enc2}\rightarrow 0$ as $n\rightarrow \infty$.

2) Error occurs at decoder 1 if \\

$E_3$: The transmitted
$(\ubf_{10}(1),\vbf_{20}(1,k_{20},1),\vbf_{11}(1,k_{11},1,k_{20},1),\ubf_{11}(1,1))$
do not satisfy (\ref{event:joint typical at dec1}).

$E_4$:
$(\ubf_{10}(j_{10}),\vbf_{20}(j_{20},k_{20}^{'},j_{10}),\vbf_{11}(j_{11},k_{11}^{'},j_{20},k_{20}^{'},j_{10}),\ubf_{11}(j_{11},j_{10}))$
satisfy (\ref{event:joint typical at dec1})

\hspace{.6cm} where $(j_{11},j_{10})\neq (1,1)$.\\

Let $A(j_{10},j_{20},k_{20}^{'},j_{11},k_{11}^{'})$ denote the
event $E_4$. We have \bqa P_{e,dec1}&\leq& P\{E_3\cup E_4\}\leq
P(E_3)+P(E_4)\\
&\leq& \epsilon + \sum_{(j_{10}j_{20}j_{11})\neq (111),k_{20}^{'},k_{11}^{'}}P(A(j_{10},j_{20},k_{20}^{'},j_{11},k_{11}^{'}))\\
&\leq& \epsilon + \sum_{j_{10}\neq
1,j_{20},k_{20}^{'},j_{11},k_{11}^{'}}P(A(j_{10},j_{20},k_{20}^{'},j_{11},k_{11}^{'}))\\
&+& \sum_{j_{20}\neq 1,k_{20}^{'},j_{11}\neq
1,k_{11}^{'}}P(A(1,j_{20},k_{20}^{'},j_{11},k_{11}^{'})) +
\sum_{j_{11}\neq
1,k_{11}^{'}}P(A(1,1,k_{20}^{'},j_{11},k_{11}^{'})) \eqa Take
$P(A(1,1,k_{20}^{'},j_{11},k_{11}^{'}))$ for example,
\bqa && P(A(1,1,k_{20}^{'},j_{11},k_{11}^{'}))\\
&=&\sum_{(\vbf_{11},\vbf_{20},\ubf_{11},\ubf_{10},\ybf_1)\in
T_{\epsilon}}
p(\ubf_{10}\vbf_{20})p(\ubf_{11}|\ubf_{10})p(\vbf_{11}|\vbf_{20}\ubf_{10})p(\ybf_1|\ubf_{10}\vbf_{20})\\
&\leq&
2^{-n[H(U_{10}V_{20})+H(U_{11}|U_{10})+H(V_{11}|V_{20}U_{10})+H(Y_1|V_{20}U_{10})-H(U_{10}V_{20}V_{11}U_{11}Y_1)-\epsilon]}\\
&\leq&
2^{-n[I(U_{11};Y_1V_{20}|U_{10})+I(V_{11};Y_1U_{11}|V_{20}U_{10})]}
\eqa

So, \bqa P_{e,dec1}&\leq& \epsilon +
2^{-n[I(V_{11}U_{11}V_{20}U_{10};Y_1)-(L_{11}+L_{20}+R_{10})-\epsilon]}\\
&+&
2^{-n[I(V_{11}U_{11}V_{20};Y_1|U_{10})-(L_{11}+L_{20})-\epsilon]}
+2^{-n[I(U_{11};Y_1V_{20}|U_{10})+I(V_{11};Y_1U_{11}|U_{10}V_{20})-L_{11}-\epsilon]}
\eqa $\epsilon$ can be made arbitrarily small by letting
$n\rightarrow \infty$. The conditions (\ref{eq:rate
3})-(\ref{eq:rate 5}) guarantee that $P_{e,dec1}\rightarrow 0$ as
$n\rightarrow \infty$.

3) The error at decoder 2 can be similarly analyzed and conditions
(\ref{eq:rate 6})-(\ref{eq:rate 8}) will ensure that
$P_{e,dec2}\rightarrow 0$ as $n\rightarrow \infty$.
\end{proof}

\end{appendix}

\bibliographystyle{C://localtexmf/caoyibib/IEEEbib}
\bibliography{C://localtexmf/caoyibib/Journal,C://localtexmf/caoyibib/Conf,C://localtexmf/caoyibib/Book}

\begin{thebibliography}{10}

\bibitem{Devroye:06IT}
N.~Devroye, P.~Mitran, and V.~Tarokh,
\newblock ``{Achievable rates in cognitive radio channels},''
\newblock {\em IEEE Trans. on Information Theory}, vol. 52, pp. 1813--1827, May
  2006.

\bibitem{Wu:07IT}
W.~Wu, S.~Vishwanath, and A.~Arapostathis,
\newblock ``{Capacity of a class of cognitive radio channels: interference
  channels with degraded message sets},''
\newblock {\em IEEE Trans. on Information Theory}, vol. 53, pp. 4391 -- 4399,
  Nov. 2007.

\bibitem{Jovicic:06ISIT}
A.~Jovicic and P.~Vishwanath,
\newblock ``{Cognitive radio: an information-theoretic perspective},''
\newblock in {\em Proc. IEEE ISIT'06}, Seattle, USA, Jul. 2006, pp. 2413--2417.

\bibitem{kramer:06ITA}
I.~Maric, R.~Yates, and G.~Kramer,
\newblock ``{The strong interference channel with unidirectional
  cooperation},''
\newblock in {\em Proceedings of UCSD Workshop on Information Theory and its
  Applications}, San Diego, CA, USA, Feb. 2006.

\bibitem{Gelfand&Pinsker:80PCIT}
S.~Gel'fand and M.~Pinsker,
\newblock ``{Coding for channels with random parameters},''
\newblock {\em Probl. Contrl. and Inf. Theory}, vol. 9, no. 1, pp. 19--31,
  1980.

\bibitem{Han&Kobayashi:81IT}
T.~Han and K.~Kobayashi,
\newblock ``{A new achievable rate region for the interference channel},''
\newblock {\em IEEE Trans. on Information Theory}, vol. IT-27, pp. 49--60, Jan.
  1981.

\bibitem{Sridharan&Vishwanath:07ITW}
S.~Sridharan and S.~Vishwanath,
\newblock ``{On the capacity of a class of MIMO cognitive radios},''
\newblock in {\em IEEE Information Theory Workshop (ITW 2007)}, Lake Tohoe, CA,
  USA, Sept. 2007.

\bibitem{Jiang&Xin:08IT}
J.~Jiang and Y.~Xin,
\newblock ``On the achievable rate regions for interference channels with
  degraded message sets,''
\newblock {\em IEEE Transactions on Information Theory}, vol. Vol. 54, no. 10,
  pp. 4707--4712, Oct. 2008.

\bibitem{Maric:08ETT}
I.~Maric, A.~Goldsmith, G.~Kramer, and S.~Shamai(Shitz),
\newblock ``On the capacity of interference channels with one cooperating
  transmitter,''
\newblock {\em European Transactions on Telecommunications}, vol. 19, pp.
  405--420, April 2008.

\bibitem{Marton:79IT}
K.~Marton,
\newblock ``{A coding theorem for the discrete memoryless broadcast channel},''
\newblock {\em IEEE Trans. Inform. Theory}, vol. 25, no. 1, pp. 306--311, May
  1979.

\bibitem{Tuninetti:07ITA}
D.~Tuninetti,
\newblock ``{On interference channels with generalized feedback (IFC-GF)},''
\newblock in {\em Proceedings of ITA'07 Workshop}, UCSD, Feb. 2007.

\bibitem{Cao&Chen:07ISIT}
Y.~Cao and B.~Chen,
\newblock ``{An achievable rate region for interference channels with
  conferencing},''
\newblock in {\em Proc. IEEE Int. Symp. Information Theory (ISIT)}, Nice,
  France, June 2007.

\bibitem{Cao&Chen:08Asilomar}
Y.~Cao and B.~Chen,
\newblock ``{Interference channel with one cognitive transmitter},''
\newblock in {\em Proc. IEEE Asilomar Conference on Signals, Systems and
  Computers}, Pacific Grove, CA, Oct. 2008.

\bibitem{ElGamal&VanderMeulen:81IT}
A.~El Gamal and E.~C. van~der Meulen,
\newblock ``A proof of marton's coding theorem for the discrete memoryless
  broadcast channel,''
\newblock {\em IEEE Transactions on Information Theory}, vol. Vol.27, no. 1,
  pp. 120--122, Jan. 1981.

\bibitem{Csiszar&Korner:78IT}
I.~Csisz\'{a}r and J.~K\"{o}rner,
\newblock ``{Broadcast channels with confidential messages},''
\newblock {\em IEEE Trans. Inform. Theory}, vol. 24, no. 3, pp. 339--348, May
  1978.

\bibitem{Gelfand&Pinsker:80PIT}
S.~I. Gel'fand and M.~S. Pinsker,
\newblock ``{Capacity of a broadcast channel with one deterministic
  component},''
\newblock {\em Probl. Inform. Transm.}, vol. 16, no. 1, pp. 17--25, Jan.-Mar.
  1980.

\bibitem{Carleial:75IT}
A.~B. Carleial,
\newblock ``{A case where interference does not reduce capacity},''
\newblock {\em IEEE Trans. Inform. theory}, vol. 21, pp. 569--570, Sep. 1975.

\bibitem{Willems:85IT}
F.~M.~J. Willems,
\newblock ``{The discrete memoryless multiple-access channel with cribbing
  encoders},''
\newblock {\em IEEE Trans. on Inf. Theory}, vol. 31, pp. 313--327, May 1985.

\bibitem{Costa:83IT}
M.~Costa,
\newblock ``{Writing on dirty paper},''
\newblock {\em IEEE Trans. Inform. Theory}, vol. 29, pp. 439--441, May 1983.

\bibitem{Cover&Elgamal:79IT}
T.~M. Cover and A.~El Gamal,
\newblock ``{Capacity theorems for the relay channel},''
\newblock {\em IEEE Trans. Inform. Theory}, vol. 25, pp. 572--584, 1979.

\bibitem{Costa:85IT}
M.~Costa,
\newblock ``{On the Gaussian interference channel},''
\newblock {\em IEEE Trans. Inform. Theory}, vol. 31, pp. 607--615, Sept. 1985.

\bibitem{Sason:04IT}
I.~Sason,
\newblock ``{On achievable rate regions for the Gaussian interference
  channel},''
\newblock {\em IEEE Trans. Inform. theory}, vol. 50, pp. 1345--1356, Jun. 2004.

\bibitem{Thomas:87IT}
J.~A. Thomas,
\newblock ``Feedback can at most double gaussian multiple access channel
  capacity,''
\newblock {\em IEEE Transactions on Information Theory}, vol. Vol.33, pp.
  711--716, Sept. 1987.

\end{thebibliography}

  \end{document}